\documentclass[aps,prb,twocolumn,showpacs]{revtex4}

\usepackage{graphicx}

\begin{document}

\title{Geometry and the Hidden  Order of Luttinger Liquids: the
Universality of Squeezed Space.}

\author {H.V. Kruis}
\author{I.P. McCulloch}
\email{ianmcc@lorentz.leidenuniv.nl}
\author{Z. Nussinov}
\altaffiliation{Present address: Theoretical Division,
Los Alamos National Laboratory, Los Alamos, NM 87545, USA.}
\email{zohar@viking.lanl.gov}
\author{J. Zaanen}
\email{jan@lorentz.leidenuniv.nl}
\affiliation{Instituut Lorentz for Theoretical Physics, 
Leiden University, P.O. Box 9506, NL-2300 RA Leiden, 
The Netherlands}


\begin{abstract}
We present the case that Luttinger liquids are characterized
by a form of hidden order which is similar, but distinct
in some crucial regards, to the hidden order characterizing 
spin-1 Heisenberg chains. We construct a string correlator
for the Luttinger liquid which is similar to the string
correlator constructed by den Nijs and Rommelse for the
spin chain. We re-analyze the spin one chain, introducing
a precise formulation of the geometrical principle behind
the so-called `squeezed space' construction, to demonstrate 
that the physics at long wavelength can be reformulated in
terms of a $Z_2$ gauge theory. Peculiarly, the normal spin chain
lives at infinite gauge coupling where it is characterized
by deconfinement. We identify the microscopic conditions required
for confinement thereby identifying a novel phase of the spin-chain.
We demonstrate that the Luttinger liquid can be approached in
the same general framework. The difference from the spin chain is
that the gauge sector is critical in the sense that the Luttinger
liquid is at the phase boundary where the $Z_2$ local symmetry emerges.
In addition, the `matter' (spin) sector is also critical. 
We evaluate the string correlator analytically for the
strongly coupled Hubbard model and further
we demonstrate that the squeezed
space structure is still present even in the non-interacting fermion gas.
This adds new insights to the meaning of bosonization. These structures
are hard-wired in the mathematical structure of bosonization and this
becomes obvious by considering string correlators. Numerical results
are presented for the string correlator using a non-abelian version
of the density matrix renormalization group algorithm, confirming in
detail the expectations following from the theory. We conclude with
some observations regarding the generalization of bosonization to higher
dimensions.
\end{abstract}

\maketitle

% some conveniences

\newcommand{\Ostr}{O_{str}}
\newcommand{\Kc}{{K_c}}
\newcommand{\Ks}{{K_c}}

% horizontal size to use for figures
\newcommand{\figurewidth}{0.8\hsize}

% command to display a rotated figure scaled to width \figuresize
\newcommand{\showgrace}[1]
{
	\vspace{2mm}
	\includegraphics[angle=270,width=\figurewidth]{#1}
	\vspace{5mm}
}

\section{Introduction}
\label{sec:intro}

The Luttinger liquid, the metallic state of one dimensional electron
matter, is an old subject which is believed to be fully understood.
In the 1970's the bosonization theory was developed which has a similar
status as the Fermi-liquid theory, making possible to compute long wavelength
properties in detail with only a small number of input 
parameters\cite{VoitSchulz,Stone}. 
In the present era, the theory is taken for granted,
and it has found many applications, most recently in the
context of nanophysics\cite{Dekker}. Here
we will attempt to persuade the reader that there is still 
something to be learned about the fundamentals of the Luttinger
liquid. 

In first instance, it is intended as a clarification of some 
features of the Luttinger liquid which appear as rather mysterious in
the textbook treatments. We make the case that a physical 
conception is hidden in the mathematics of the standard
treatise. This physical conception might alternatively be
called `hidden order', `critical gauge deconfinement' or
`fluctuating bipartite geometry'. It all refers to the same
entity, viewed from different angles. This connection was 
first explored in our previous letter\cite{Letter}, here we
expand on these ideas to
%This improved understanding does 
yield some practical consequences:
(a) we identify symmetry principles allowing a sharp distinction 
between Luttinger liquids and for instance the bosonic liquids 
found in spin-1 chains\cite{Nijs} (the `no-confinement' principle, sections 
\ref{sec:gauge}, \ref{sec:luttinger}), 
(b) we identify a new competitor of the Luttinger liquid
(the manifestly gauge invariant superconductor, section \ref{sec:luttinger}, a
close sibling of the superfluid $t-J$ model of Batista and 
Ortiz\cite{Batista}), and
(c) these insights go hand in hand with  special `string' (or `topological')
correlation functions which makes possible unprecedented precision
tests of the analytical theory by computer simulations, offering also
advantages for the numerical determination of exponents (section \ref{sec:numerical}). 

This pursuit was born out from a state of confusion we found ourselves
in some time ago, caused by a view on the Luttinger liquid from an 
unusual angle. Our interest was primarily in what is now 
called `stripe 
fractionalization'\cite{Zaanenphilmag,Nussinov,Sachdev0,Sachdev1}. 
Stripes refer to textures found
in doped Mott-insulators in higher dimensions. These can be alternatively
called `charged domain walls'\cite{Zaanen89}: 
the excess charges condense on $d-1$
dimensional manifolds, being domain walls in the colinear
antiferromagnet found in the Mott-insulating domains separating the
stripes. Evidence accumulated that such a stripe phase
might be in close competition with the high-$Tc$ superconducting state
of the cuprates\cite{Zaanenphilmag,Kivelson} 
and this triggered a theoretical effort aimed at an
understanding of stripe quantum liquids. The idea emerged that in
principle a superconductor could exist characterized by 
quantum-delocalized stripes which are however still forming intact
domain walls in the spin system. Using very similar arguments as
found in sections \ref{sec:gauge} and \ref{sec:luttinger} 
of this paper, it can then be argued
that several new phases of matter exist governed by Ising gauge
theory. This is not the subject of this paper and we refer the
interested reader to the 
literature\cite{Zaanenphilmag,Nussinov,Sachdev0,Sachdev1}. 
However, we realized early
on that these ideas do have an intriguing relationship with
one dimensional physics.

Specifically, we were intrigued by two results which, although well 
known, do not seamlessly fit into the 
Luttinger liquid mainstream: (a) the hidden order
in Haldane spin chains as discovered by den Nijs and Rommelse\cite{Nijs}, 
(b) the squeezed space construction as deduced by Woynarovich\cite{Woyn}, 
Ogata and Shiba\cite{ogashi} from
the $U \rightarrow \infty$ Bethe Ansatz solution of the Hubbard model.
As we will discuss in much more detail, after some further thought
one discovers that both refer to precisely the same underlying 
structure. This structure can be viewed from different sides.
Ogata and Shiba\cite{ogashi} emphasize the 
geometrical side: it can be literally
viewed as a dynamically generated `fluctuating geometry', although one
of a very simple kind. Den Nijs and Rommelse approached it using
the language of order\cite{Nijs}: 
a correlation function can be devised approaching
a constant value at infinity, signaling symmetry breaking. The analogy
with stripe fractionalization makes it clear that it can also be characterized
as a deconfinement phenomenon in the language of gauge theory.

Whatever way one calls it, this refers to a highly organized, dynamically
generated entity. The reason we got confused is that there is no mention
whatsoever in the core literature of the Luttinger liquid of how these
squeezed spaces etcetera fit in the standard bosonization lore. To
shed some light on these matters we found inspiration in the combined 
insights of den Nijs-Rommelse and Ogata-Shiba and we constructed
a den Nijs type `string' correlator but now aimed at the detection of the
squeezed space of Ogata and Shiba. This is the principal device that we use, 
and it has the form,
\begin{equation}
\Ostr ( | i - j| ) =
\langle \; \vec{S}_i \; \left[ \Pi_{l=i}^j (-1)^{n_i}
\right] \;  \vec{S}_j \; \rangle \; ,  
\label{otop0}
\end{equation}
where $\vec{S}_i$ is the spin-operator on site $i$ while $n_l$ measures
the charge density. By studying the behavior of this correlator one can 
unambiguously establish the presence of squeezed-space like structures. 
We spend roughly the first half of this paper explaining how
this works and what it all means. In section \ref{sec:gauge} we start with
a short review of the den Nijs-Rommelse work on the $S=1$ 
`Haldane'\cite{Haldane}
spin chains. This is an ideal setting to develop the conceptual
framework. We subsequently reformulate the spin chain `string'
correlator in a geometrical setting which makes the relationship
with the Woynarovich-Ogata-Shiba squeezed space manifest. We finish
this section with the argument why it is Ising gauge theory in disguise.
This is helpful, because the gauge theory sheds light on the limitations
of the squeezed space: we present a recipe of how to destroy the squeezed
space structure of the spin chain. 

In section \ref{sec:luttinger} we revisit Woynarovich-Ogata-Shiba. 
The string correlator Eq. (\ref{otop0})
is formulated and subsequently investigated in the large $U$ limit. This
analysis shows that the Luttinger liquid (at least for large $U$) can be 
viewed as the critical version of the Haldane spin chain. It resides at
the phase transition where the gauge invariance emerges, while the matter
fields are critical as well. In this section we also argue why
the squeezed space of the {\em electron} liquid {\em cannot} be destroyed.
This turns out to be an unexpected consequence of Fermi-Dirac statistics.

In the remaining two sections the string correlator is used to interrogate
the Luttinger liquid regarding squeezed space away from the strong
coupling limit. In section \ref{sec:NonInteracting} we demonstrate in a few lines a most
surprising result: squeezed space exists even in the non-interacting
spinful fermion gas! This confirms in a dramatic way that squeezed 
space is deeply rooted in fermion statistics; it is a complexity
price one has to pay when one wants to represent fermion dynamics 
in one dimension in terms of bosonic variables.

In section \ref{sec:bosonization} we turn to bosonization. Viewing the bosonization formalism
from the perspective developed in the previous sections it becomes
clear that the squeezed space structure is automatically wired into
the structure of the theory. In this regard, the structure of bosonization 
closely parallels the exact derivations presented in section \ref{sec:luttinger}. 
In section
\ref{sec:numerical} we present numerical density matrix renormalization group (DMRG) 
calculations for the string correlators starting from the Hubbard model at
arbitrary fillings and interaction strength, employing a non-abelian 
algorithm. These results confirm in a great detail the expectations 
build up in the previous sections: the strongly interacting limit
and the non-interacting gas are smoothly connected and in the scaling
limit the string correlator Eq. (\ref{otop0}) isolates the spin only
dynamics regardless the microscopic conditions. This has also practical
consequences; we deliver the proof of principle that the non-universal
exponents associated with the logarithmic corrections showing up in
the spin-correlations can be addressed away from half-filling.
From the combination
of bosonization and the exact results for strong coupling suggest
that the two point spin correlator can always be written in the scaling
limit as the product of Eq(\ref{otop0}) and a charge-like
string correlator,
\begin{equation}
\langle \vec{S} (x) \vec{S} (0) \rangle \; \sim \; D_{nn} ( x ) \; \Ostr (x) \; ,
\label{twpoSS}
\end{equation}
where the `charge' string operator is defined as,
\begin{equation}
D_{nn} ( | i - j | ) \; \equiv \; \langle \; n_s (i) \;
\left[ \Pi_{l=i}^j (-1)^{n_s (l)} \right] \; n_s (j) \rangle \,
\label{Dnndef}
\end{equation}
where $n_s (i)$ is 1 for a {\em singly} occupied site and  0 otherwise.
We confirm numerically that except for a non-universal amplitude the
relation Eq. (\ref{twpoSS}) seems always satisfied at long distances. 

The conclusion
to this paper addresses the broader perspective including
the relation to stripe fractionalization in 2+1 dimensions.                

\section{Geometry, gauge theory and Haldane spin chains.}
\label{sec:gauge}

The `Haldane'\cite{Haldane} $S=1$ Heisenberg spin chains are an ideal stage
to introduce the notions of `hidden order', squeezed space,
and the relation with Ising gauge theory.   
 These systems are purely bosonic, 
i.e. dualization is not required for the identification
of  the bosonic fields,
and the powers of bosonic field theory can be utilized with great
success to enumerate the physics completely. We refer in particular 
to the mapping by den Nijs and Rommelse\cite{Nijs} on surface statistical
physics. We are under the impression that this way of thinking
is not widely disseminated and we start out reviewing some
highlights. In the surface language, the meaning of `hidden-order' becomes
particularly simple (section \ref{subsec:haldane}). We subsequently use these
simple insights to reformulate this hidden order in the geometrical
language, the `squeezed space' (section \ref{subsec:squeezed}). 
The next benefit of the Haldane  chain is that the identification
of squeezed space geometry with Ising gauge theory is literal
(section \ref{subsec:ising}). This sets the conceptual framework within which we view
the Luttinger liquid.

\subsection{Haldane spin chains: a short review}
\label{subsec:haldane}

Let us first review some established wisdoms concerning the Haldane
spin chains. The relevant model is a  standard Heisenberg model
for $S=1$ extended by biquadratic exchange interactions and single-ion
anisotropy,
\begin{equation}
H = \sum_{<ij>} \vec{S}_i \cdot \vec{S}_j + \alpha   \sum_{<ij>} (
\vec{S}_i \cdot \vec{S}_j )^2 + D \sum_i (S^z_i)^2 \; .
\label{haldanespchain}
\end{equation}
In the proximity of the Heisenberg point ($\alpha = D = 0$) the ground state
is a singlet, separated by a finite energy gap from propagating triplet
excitations. It was originally believed that the long distance physics
was described by an $O(3)$ non-linear sigma model\cite{Haldane}, 
suggesting that the ground state is featureless singlet. 
However, Affleck et al\cite{AKLT} discovered
that for $\alpha = 1/3, D=0$ the exact ground state wavefunction can be
deduced, having a particularly simple form. This `AKLT' wavefunction can
be parametrized as follows. Split the $S = 1$ microscopic singlets into
two Schwinger bosons $ |S =1, M_s \rangle \sim b^{\dagger}_{i 1, \alpha} 
b^{\dagger}_{i 2, \alpha}$. The individual Schwinger bosons carry $S=1/2$
and the wavefunction is constructed by pairing, say, the `1' boson with
a `1' boson on the left neighboring site forming a singlet of valence
bond, and the same with the '2' boson with its counterpart on the right
neighbor,
\begin{widetext}
\begin{equation}
\begin{array}{rcl}
| \Psi \rangle_{AKLT} & = & 
2^{-N/2} \left[ \cdots (b^{\dagger}_{i-1; 1\uparrow}
b^{\dagger}_{i; 1\downarrow} - b^{\dagger}_{i-1; 1\downarrow}
b^{\dagger}_{i; 1\uparrow} ) \;
(b^{\dagger}_{i; 2\uparrow} b^{\dagger}_{i+1; 2\downarrow} - 
b^{\dagger}_{i; 2\downarrow} b^{\dagger}_{i+1; 2\uparrow} ) 
 \right. \\ & & \left. \times
 (b^{\dagger}_{i+1; 1\uparrow}
b^{\dagger}_{i+2; 1\downarrow} - b^{\dagger}_{i+1; 1\downarrow}
b^{\dagger}_{i+2; 1\uparrow} ) \cdots \right] | \textrm{vac} \rangle
\end{array}
\label{AKLTpsi}
\end{equation}
\end{widetext}

This wave function clearly has to do with a translation symmetry 
breaking involving nearest-neighbor singlet pairs, although in terms
of spin-degrees of freedom which are different from the elementary
spins. It has become a habit to call it `valence bond solid order',
i.e. to link it exclusively to the tendency in the spin system to
form spin $1/2$ singlet pairs. 
Den Nijs and Rommelse\cite{Nijs} added a deep understanding of
the physics of these bosonic spin chains by introducing the mapping
on surface statistical physics. Although the AKLT wave function
is a correct prototype for the ground state of the Heisenberg chain,
it is not helpful with regard to what else can happen. On the other
hand, by employing the formidable powers of surface statistical
physics there are no secrets and it yields a natural view
on the physics of the spin-chains. A highlight is
their demonstration that this vacuum can be understood by a
a non-local (`topological') order parameter structure in terms of
the real $S=1$ spin degrees of freedom.  The measure
of order is the asymptotic constancy of a correlation function.
The conventional two-point spin correlator in the Haldane chain
decays exponentially for large $|i-j|$,
\begin{equation}
\langle S^z_i \;  S^z_j \rangle \sim e^{- | i - j |  / \xi}
\label{2ptspha}
\end{equation}

However, considering the following non-local spin correlator
(or 'string' correlator),
\begin{equation}
\langle S^z_i \; \left[ \Pi_{l=i}^j (-1)^{S^z_l} \right] \; S^z_j \rangle
\sim \text{constant}
\label{dNstring}
\end{equation}
signaling a form of long range order which only becomes visible when
probed through the non-local correlator Eq. (\ref{dNstring}). For
this reason it was called `hidden order'. A main purpose of this section 
will be to introduce a more precise definition of this order.

Den Nijs and Rommelse\cite{Nijs}  deduced the string correlator using 
the insights following from the path-integral mapping onto surface 
statistical physics. A first, crucial observation is that the natural
basis for the spin chain is not in terms of generalized coherent states,
but instead simply in terms
of the microscopic $M_s = 0, \pm 1$ states of individual spins.
Marshall signs are absent and these states can be parametrized in terms
of flavored bosons $b^{\dagger}_{0}, b^{\dagger}_{\pm 1}$ subjected to
a local constraint $\sum_{\alpha} b^{\dagger}_{\alpha} b_{\alpha} = 1$.
The spin operators become,
\begin{eqnarray}
S^z_i & = & n_{i,1} - n_{i, -1} \nonumber \\
S^+_i & = & \sqrt{2} \left(  b^{\dagger}_{i, 1} b^{}_{i, 0 } +
 b^{\dagger}_{i, 0} b^{}_{i, -1 } \right) \nonumber \\
S^-_i & = & \sqrt{2} \left(  b^{\dagger}_{i, 0} b^{}_{i, 1 } +
 b^{\dagger}_{i, -1} b^{}_{i, 0 } \right)
\label{spinsMs}
\end{eqnarray}

A second crucial observation is that because of the constraint the
problem is isomorphic to that of a (directed) quantum string living
on a square lattice. This is somewhat implicit in the original
formulation by den Nijs and Rommelse, but used to great effect by
Eskes {\em et al.}\cite{Eskes}. The mapping is elementary. A lattice string
corresponds with a connected trajectory of `particles' living on
a lattice and this string
can in turn be parametrized by a center of mass coordinate and the
set of links connecting all particles. Consider only `forward moving'
links (the string is directed) and identify a nearest-neighbor link
with a $M_s = 0$ bond, and an `upward' and `downward' next-nearest-neighbor
link (`kinks')  with $M_s = 1$ and $M_s = -1$ states of the spin on a site 
of the Haldane chain, respectively. It is easy to convince oneself that 
every state 
in the Hilbert space of the spin chain corresponds with a particular
string configuration. The XY terms are responsible for the creation
of kink-antikink pairs and the propagation of individual kinks along
the string, while Ising terms govern the interactions between 
the kinks. In the path integral formulation, quantum strings spread out
into world-sheets and the world-sheet of the lattice string corresponds
with a surface statistical physics which is completely understood:
the restricted solid-on-solid (RSOS) surface. 

\begin{figure}[tbh]
\includegraphics[width=\figurewidth]{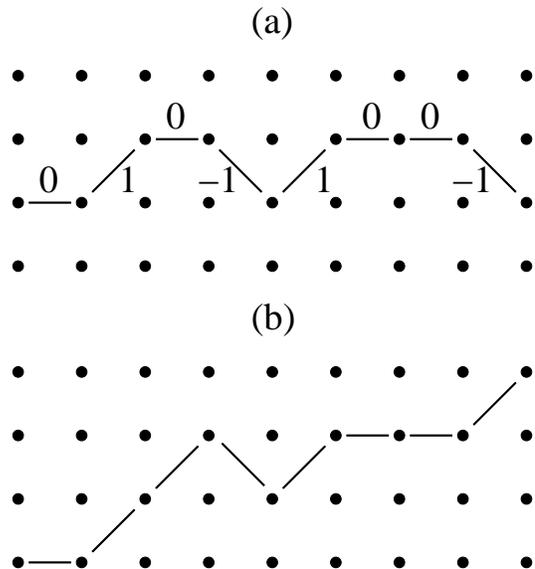}
\caption{
Mapping of the spin chain on a directed quantum string living
on a lattice \protect\cite{Eskes}.
The $M_s = 0, \pm 1$ states of the spin
chain are equivalent to horizontal-, and upward/downward diagonal
links tracing out the trajectory of the string on the lattice. The
$XY$ terms in the spin  Hamiltonian correspond with the kinetic energy
of the string problem causing both the creation of kink-antikink pairs
(the $\pm 1$ bonds) and the coherent propagation of individual kinks.
At the Heisenberg- and AKLT points hidden
order is present. Although kinks are proliferated their direction
is ordered: every up kink is followed by a down kink (a). In the 
string representation this just means that the string pins to the
links of the lattice.
In the rough ($XY$) phase kinks have proliferated while their direction
is disordered as well (b).}
\label{eskesstr}
\end{figure}

The topological order of the Haldane spin chain translates into a
simple form of order in the surface language: the disordered flat
phase. The $+1$ and $-1$ kinks on the time slice turn into up and
down steps on the world-sheet in space time, see Fig. (\ref{eskesstr}). 
In the disordered flat
phase these steps have proliferated (kinks occur at finite
density while they are delocalized) but on this surface 
every `up' step is followed by a down step and the surface
as a whole is still flat, pinned to the lattice. In the string
language the order is therefore manifest, but it becomes elusive
when translated back to the spin system. It implies that the
ground state of the spin chain is a coherent superposition
of a special class of states.
These are composed of indeterminate mix of $0, \pm 1$ states.
Take the $0$'s as a reference vacuum and view 
the $M_s = \pm 1$ states as particles carrying an in internal `flavor'
$\pm 1$. These particles are delocalized.
However, every $+1$ particle is followed by a $-1$ particle,
modulo local violations (virtual excitations) which can be integrated out
perturbatively (fig. \ref{eskesstr}). The hidden order is thereby nothing
else than the staggered order of the $\pm 1$ flavors of the `spin particles'.
This order is not seen by the spin-1 operators because these also
pick up the {\em positional} disorder of the $\pm 1$ `particles'.

\subsection{Squeezed space: sublattice parity as a gauge freedom.}
\label{subsec:squeezed}

\begin{figure}[tbh]
\includegraphics[width=\figurewidth]{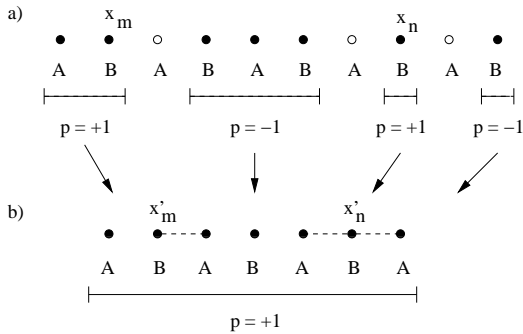}
\caption{The geometrical mapping from `full' (a) to `squeezed' space (b).
Given that some antiferromagnet lives in squeezed space, all that matters
is the fate of the sublattice parity $p$. When sites are reinserted,
the sublattice parity of the system in squeezed space flips every time
a hole is passed when viewed in full space. These sublattice parity flips 
hide the order present in squeezed space.}
\label{Squeezepict}
\end{figure}

String correlators of the kind  Eq. (\ref{dNstring}) have the purpose 
of `dividing out' the positional disorder with the effect that the
order of the `internal' $\pm 1$ flavors becomes observable. In order
to see the similarity with the phenomena occurring in the Luttinger
liquid we need a more precise description of how this `division'
is accomplished than that found in the original literature. It 
amounts to a geometrical mapping of a simple kind.
The string correlator can be written in terms of the bosons as, 
\begin{equation}
\begin{array}{rcl}
\langle S^z_i \left[ \Pi_{l=i}^j (-1)^{S^z_l} \right] S^z_j \rangle
& \equiv &
\left\langle ( n_{i,1} - n_{i, -1})
\right. \\ & & \times \;
\Pi_{l=i}^j (-1)^{1 - n_{l, 0}}
\\ & & \times \; \left.
( n_{j,1} - n_{j, -1})
\right\rangle
\end{array}
\label{bostring}
\end{equation}
Why is this tending to a constant while the two-point spin correlator
is decaying exponentially? From the discussion in section \ref{subsec:haldane} it follows
that modulo local fluctuations the ground state
wave function has the form,
\begin{widetext}
\begin{equation}
|\Psi \rangle = \sum \; a (x_1, x_2, ..., x_{2i}, x_{2i+1}, ...) \;
| x_1 (1), x_2 (-1), ..., x_{2i} (-1), x_{2i+1} (1), ... \rangle  
\label{halwave}
\end{equation}
\end{widetext}
where the $x_i$'s refer to the positions of the $\pm 1$ particles
on the chain, and the amplitudes $a$ are independent from the
`internal' ($\pm 1$)  degrees of freedom; these `internal' Ising spins
show the antiferromagnetic order. In order to construct a two point correlator
capable of probing this `internal' order it is necessary to redefine the
space in which the internal degrees of freedom live.  Start out with the 
full spin chain and for each configuration, whenever a site
occupied by a 0-particle is found remove this site and shift, say,
all right neighbors to the left, see fig. (\ref{Squeezepict}). 
This new space is called
`squeezed space' and the effect of the map from `full' to squeezed space
is such that every configuration appearing in Eq. (\ref{halwave}) maps on the
same antiferromagnetic order as realized on the squeezed chain.
 
Obviously,
if it would be possible to probe squeezed space directly, the hidden order
would be measurable using conventional two point correlators.
The string correlator achieves just this purpose.
All that matters is that the order in squeezed space is a staggered
(antiferromagnetic) order. For such order one needs a bipartite geometry:
it should be possible to divide the lattice into $A$ and $B$ sublattices
such that every site on the $A$ sublattice is neighbored by $B$
sublattice sites and vice versa. In one dimension any space is bipartite
(even the continuum). This subdivision can be done in two ways: 
$\cdots A - B - A - B \cdots$ or  $\cdots B - A - B - A \cdots$, corresponding
with the $Z_2$ valued quantity we call sublattice parity. For a normal
lattice the choice of sublattice parity is arbitrary, it is a `pure gauge'.
However, in the mapping of squeezed to full space it becomes `alive', actually
in a way which is in close correspondence to the workings of a dynamical
$Z_2$ gauge field as will become clear later. 
Consider what happens when squeezed space is unsqueezed
(Fig.\ref{Squeezepict}). When a `0' particle including its site is reinserted,
the `flavor' site, say, on its right side is 
shifted one lattice constant to the right. The
effect is that relative to the reference sublattice parity of squeezed
space the sublattice parity in unsqueezed space changes sign every time
a `0' particle is passed. The effect is that flips in the sublattice
appear to be `bound' to the $0$ particles viewing matters in full space.
In order to interrogate the `flavor'  order in squeezed space one has to
remove these sublattice parity flips. This can be achieved by
multiplying the spin with a minus one every time a 0 particle is encountered:
$(-1) \times (-1)^{l+1} = (-1)^l$. The den Nijs string operator is constructed
to precisely achieve this purpose,
%\begin{widetext}
\begin{equation}
\begin{array}{l}
\langle ( n_{i,1} - n_{i, -1}) 
\left[ \Pi_{l=i}^j (-1)^{1- n_{l, 0}} \right] ( n_{j,1} - n_{j, -1})
\rangle
\\ \quad
\equiv 
\langle (-1)^i ( n_{i,1} - n_{i, -1}) \left[ 
\Pi_{l=i}^j (-1)^{n_{l, 0}} \right]
(-1)^j ( n_{j,1} - n_{j, -1}) \rangle
\label{strstag}
\end{array}
\end{equation}
%\end{widetext}
Hence the string correlator measures the spin order in squeezed space
by removing the sublattice parity flips. The positional disorder of
the particles is equivalent to motions of the sublattice
parity flips, scrambling the order living in squeezed space, and these
are removed by the string operators.  

\begin{figure}[tbh]
\includegraphics[width=\figurewidth]{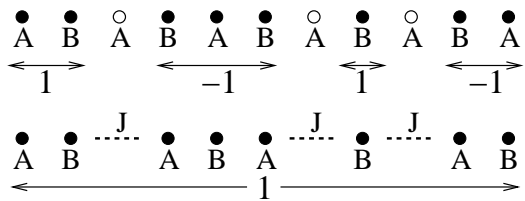}
\caption{Same as Fig. (\ref{Squeezepict}) but now for the case
that an Ising long range order is present in squeezed space,
corresponding with the hidden order of the $S=1$ Heisenberg
spin chain. From such pictures one can directly deduce the
workings of string operators by just focusing on the spins.
However, matters are equally meaningful in the absence
of long range order in squeeze space, and the formulation
terms of a geometrical mapping involving sublattice
parity is more general.}
\label{Squspins}
\end{figure}

The above argument emphasizes the geometrical nature of the mechanism
hiding the order. It might at this point appear as a detour because 
one arrives at the same conclusion  by just focusing on the `flavor' 
orientations, see Fig. (\ref{Squspins}). However,
as will become clear in later sections, the construction is still
applicable even when the spin system in squeezed space is disordered.
Hence, it is more general and rigorous to invoke the geometrical sublattice 
parity as a separate degree of freedom in addition to the degrees of freedom 
populating squeezed space.

\subsection{Squeezed spaces and Ising gauge theory.}
\label{subsec:ising}

At first sight, it might appear that sublattice parity is not quite
like a normal dynamical degree of freedom. However, it is easily
seen that it is nothing else than an uncommon ultraviolet 
regularization of $Z_2$ gauge fields. From the above
discussion it is clear that the `flavor' degrees of freedom of
the $\pm 1$ particles can be regarded as independent from their
positions in unsqueezed space. These flavors are $Z_2$ valued and can be
measured by,
\begin{equation}
\tau^z_i = ( 1 - b^{\dagger}_{i,0} b^{}_{i,0} ) (-1)^i S^z_i
\label{tauop}
\end{equation}
The positions of the particles drive the uncertainty 
in the value of the sublattice parity and these are captured by
the $Z_2$ valued operators,
\begin{equation}
\sigma^z_l = (-1)^{n_{l,0}}
\label{sigmaop}
\end{equation}
and it follows that modulo a factor of order 1,
\begin{equation}
\begin{array}{ll}
\langle ( n_{i,1} - n_{i, -1}) 
\left[ \Pi_{l=i}^j (-1)^{1 - n_{l, 0}} \right] ( n_{j,1} - n_{j, -1})
\\ \quad
\propto \langle \tau^z_i \;  \left[ \Pi_{l=i}^j \sigma_l^z  \right] \; \tau^z_j
\rangle
\end{array}
\label{gastring}
\end{equation}
and in the presence of the hidden order,
\begin{equation}
\langle \tau^z_i \tau^z_j \rangle \; \propto \; e^{ -| i - j | / \xi} \;
\langle \tau^z_i  \left[ \Pi_{l=i}^j \sigma_l^z  \right] \tau^z_j \rangle
\label{wilsonl}
\end{equation}
i.e., at distances large compared to $\xi$ the correlations between
the $\tau$ spins have disappeared but they re-emerge when the
operator  string $ \left[ \Pi_{l=i}^j \sigma_l^z \right]$ is attached to
every spin. 

This suffices to precisely specify the governing symmetry principle:
the long distance physics is
governed by a $Z_2$ gauge field (the $\sigma$'s) minimally coupled
to spin-1/2 matter (the $\tau$'s). The strings  $ \left[ \Pi_{l=i}^j 
\sigma_l^z \right]$ simply correspond with the Wilson loop associated with
the $Z_2$ gauge fields rendering the matter correlation function
gauge invariant. The two point correlator in the $\tau$'s
is violating gauge invariance and has therefore to disappear. This
gauge invariance is emerging. It is not associated with the 
microscopic spin Hamiltonian and it needs some distance $\xi$
before it can take control. Therefore, the gauge-violating
$\langle \tau^z_i \tau^z_j \rangle$ is non zero for $|i - j| < \xi$.  

This is an interesting and deep connection: the indeterminedness
of the sublattice parity in full space is just the same as invariance
under $Z_2$ gauge transformations. One can view the squeezed space construction
as an ultraviolet regularization of $Z_2$ gauge theory,
demonstrating a simple mechanism for the `making'
of gauge symmetry which is distinct from the usual mechanism invoking
local constraints (e.g., references \onlinecite{Sachdev1,Senthil}.) 

Is this yet another formal representation or 
does it reveals new physical principle? As we will now argue, the latter
is the case. Viewing it from the perspective of the gauge theory, it
becomes immediately obvious that there is yet another possible phase
of the spin chain: the {\em confining} phase of the gauge theory. 
To the best of our knowledge this phase has been overlooked because
its existence is  not particularly obvious in the spin language.

For a good tutorial in gauge theory we refer to Kogut's review\cite{Kogut}.
Focusing on the most relevant operators, the $Z_2/Z_2$ theory
can be written as,
\begin{eqnarray}
Z & = & \int {\cal D} \tau {\cal D} \sigma e^{-S} \nonumber \\
S & = & \int d^d x d\tau \left[ J \sum_{ij} \tau_i \sigma_{ij} \tau_j
+ K \sum_{plaq} \Pi_{plaq} \sigma \right]
\label{z2z2gauge}
\end{eqnarray} 
 leaving the gauge volume implicit in the measure. $\tau$ and $\sigma$
are $Z_2$ valued fields living respectively on the sites and the
links of a (hypercubic) space-time lattice. The action of the
gauge fields is governed by a plaquette action, i.e. the product of
the fields encircling every plaquette, summed over all plaquettes.
The gauge invariance corresponds with the invariance of the action
under the flip  of the signs of all the  $\sigma$'s departing from
a site $i$, accompanied by a simultaneous flip of the $\tau_i$. This
gauge invariance implies that $\tau_i = 1 \leftrightarrow -1$ and
$\langle \; \tau_i \;  \tau_j \; \rangle = 0$ 
while $ \langle \; \tau_i \; \left[ \Pi_{\Gamma} \sigma \right] \;
\tau_j \; \rangle$ can be non-zero (with $\Gamma$ a line of bonds on the 
lattice connecting $i$ and $j$ ; the Wilson loop). This is the most
general ramification of the gauge symmetry and Eq. (\ref{wilsonl})
is directly recognized. 

\begin{figure}[tbh]
\includegraphics[width=\figurewidth]{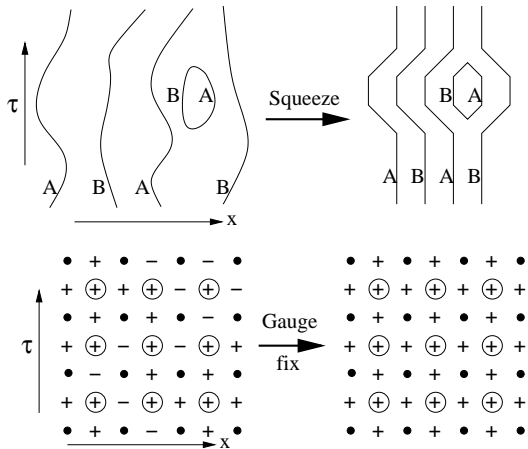}
\caption{Squeezed space mappings as geometrical interpretation of
Ising gauge theory. Although the word-lines (in space-time $x, \tau$)
 of the $\pm 1$ particles
span up a bipartite lattice for an observer which is just watching
word-lines, this bipartiteness is hidden in full space when the
particles are delocalized. This is equivalent to the conventional
lattice regularization of a $Z_2$ gauge theory involving a plaquette
action where the $+ \leftrightarrow -$ gauge invariance of the link
variables acquires the meaning that it is impossible to determine 
the bipartiteness of squeezed space by measuring in full space. The
absence of free visons (minus fluxes) does imply that the hidden bipartiteness
exists and the existence of squeezed space corresponds with deconfinement.
Taking the unitary gauge is equivalent to squeezing space.}
\label{deconf}
\end{figure}

The relation between the gauge theory and the squeezed space construction
is simple (Fig. \ref{deconf}).
The gauge invariance is just associated with the indeterminacy
of the sublattice parity in unsqueezed space. If the $0$'s would
not fluctuate one could ascribe a definite value to the sublattice
parity everywhere, and this is equivalent to choosing a unitary
gauge fix in the gauge theory. However, because of the delocalization
of the 0's one cannot say if the sublattice parity is $+1$ or $-1$ 
and this corresponds with the gauge invariance.

As is obvious from the string correlator, the $Z_2$ gauge fields
(coding for the indeterminacy of the sublattice parity) are
coupled to matter degrees of freedom being just the `flavors' living
in squeezed space. In the hidden-order/disordered flat phase these
are Ising spins showing long range order. The constancy of the
string correlator at long distances reflects this fact. From the
viewpoint of the gauge theory this appears as an 
{\em absurdity}. It means that the hidden order phase is the Higgs
phase of the $Z_2/Z_2$ gauge theory, characterized by a gauged
matter propagator becoming asymptotically constant. In the
gauge theory this can only happen in the singular limit where the
gauge coupling $K \rightarrow \infty$!

Even under the most optimal circumstances (high dimensionality), 
a Wilson loop should decay exponentially with a perimeter law
due to local fluctuations in the gauge sector. Stronger, it is
elementary that in 1+1D the Higgs and deconfining phases are
fundamentally unstable to confinement. This law can only be violated
in the singular limit $K \rightarrow \infty$. Hence, the hidden order
appears as highly unnatural within the framework of the gauge theory.
What is the reason that confinement is avoided in the Haldane spin
chain? More interestingly, what has to be done to recover the 
natural confinement state?

\begin{figure}[tbh]
\includegraphics[width=\figurewidth]{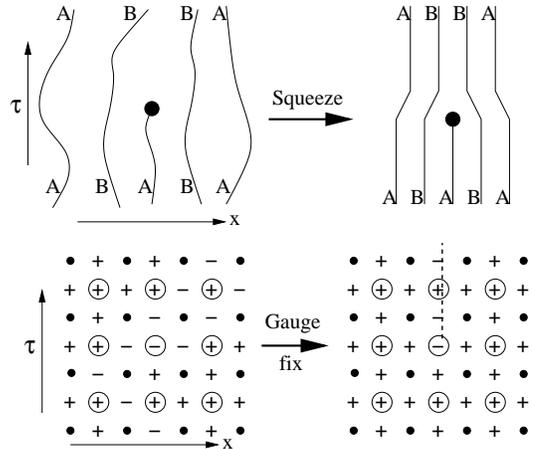}
\caption{As Fig. (\ref{deconf}) except that now a single $\pm 1$
particle is annihilated. This means that the bipartiteness of
squeezed space is destroyed and this is in one-to-one correspondence
with the presence of isolated visons (minus fluxes) in 
the $Z_2$ gauge theory formulation: the `natural' confining state.} 
\label{confi}
\end{figure}

The disorder operators in the gauge sector
are the visons or gauge fluxes. These are point-like entities
(instantons) in 1+1 dimensional space-time. For any finite 
value of the coupling constant $K$ 
these will be present at a finite density with the result that
the vacuum is confining and the implication that 
$\langle \;  \tau_i \; \left[ \Pi_{\Gamma} \sigma \right] \;
\tau_j \; \rangle \rightarrow 0 $ 
at large distances. Translating this to the geometrical
language, a vison corresponds with a process where a squeezed
space of even length on time slice $\tau$ turns into a squeezed
space of odd length on time slice  $\tau + \delta \tau$ or 
vice versa. In this way a minus gauge-flux is accumulated
on a time-like plaquette (see Fig. \ref{confi}). In terms
of the degrees of freedom of the spin chain this means that
a {\em single} $M_s = 0$ state can fluctuate into a $M_s = \pm 1$
state and vice versa. It is obvious now why the 
spin chain corresponds with the $K \rightarrow \infty$ limit
of the gauge theory, namely the Hamiltionian of the former only
contains {\em pairs} of spin raising or lowering operators
$\sim S^{+}_i S^{-}_j$. From Eq. (\ref{spinsMs}) it follows
immediately that `0' particles can only be created or annihilated
in pairs. These processes do change the length of squeezed
space but they turn even-length squeezed space into even-length
squeezed space, or odd-length squeeze space into odd-length
squeezed space. Confinement requires odd to even or even to odd
fluctuations. In the geometrical language, deconfinement means
that space-time is still bipartite although the two ways of subdividing
space-time  are indistinguishable. Confinement means that bipartiteness
is destroyed outright because squeezed space-time can no longer be
divided in two sublattices due to the presence of the visons.

Going back to the spin chain the remedy becomes obvious. First, one 
should impose a preferred direction of the spin quantization axis which can be
accomplished by taking for instance $D > 0$. Apply now a {\em transversal
field}: $B \sum_{i} S_i^x = (B/2) \sum_{i} \left[ S^+_i + S^-_i \right]$.
Given Eq. (\ref{spinsMs}), it follows that for any finite value of 
$B$ isolated 0's will turn into $\pm 1$'s and squeezed space can be 
destroyed! A somewhat delicate aspect is that the matter field is in
the fundamental representation (i.e. the $Z_2/Z_2$ matter/gauge theory).
We will analyze this elsewhere in more detail but it can be easily argued
that for any finite transverse field a featureless `Higgs-confinement'
phase will take over, characterized by exponential decay of the den Nijs
string correlator.

One can actually wonder whether under experimental circumstances the hidden
order phase can be ever truly realized in spin chains. 
The hidden order is in a sense pathological, as  deconfinement is in 1+1D:
a transverse magnetic field of any strength suffices to destroy the hidden
order. Is it at all possible to devise experiments such that transverse
fields are rigorously vanishing? Exploiting the relationship with gauge
theory, a number of other interesting conclusions can be reached regarding
of the spin chains. However, spin chains are not the real subject of this
paper, and we leave this for a future publication. The primary aim
of this section is to supply a conceptual framework for the discussion
of the more convoluted `hidden order' in the Luttinger liquids.
Let us list the important lessons to be learned from the spin chains,
and indicate how these relate to the Luttinger liquids:

\begin{itemize}

\item[1.] The central construction is squeezed space, the existence of which
can be detected using den Nijs-type string correlators.
The determination of such a correlator for the Luttinger liquid is
the subject of the next section.

\item[2.] The phases where sublattice parity flips are truly delocalized   
are characterized by an emergent $Z_2$ gauge symmetry. We make the
case that such phases can in principle occur also in the Luttinger liquid
context, while the Luttinger liquid itself resides right at the phase
boundary where the $Z_2$ local invariance emerges.

\item[3.] In the spin chains squeezed space can be destroyed by transverse
fields causing confinement. We argue that in the Luttinger liquids this
is impossible because of the fermion minus signs of the electrons, with
the ramification that squeezed space {\em is universal}.    

\end{itemize}
   
\section{Luttinger liquids: squeezed space in the large $U$  limit.}
\label{sec:luttinger}

The focus in this section
is entirely on the  Luttinger liquids which
can be regarded as continuations of those describing
the long distance physics of Hubbard models. The bottom line is
that these Hubbard-Luttinger liquids are characterized by a critical 
form of the spin-chain type hidden order as discussed in 
the previous section. This criticality has two sides:
(a) the ($Z_2$) gauge fields are critical, in the sense that the
Luttinger liquid is associated with the phase transition where 
the local symmetry emerges, (b) the matter fields (spins)
are also in a critical phase.

The argument rests again on the squeezed space construction,
and this should not come as a surprise to the reader who is
familiar with the one dimensional literature. This construction
was introduced first by Ogata and Shiba\cite{ogashi}, who rediscovered
earlier work by Woynarovich\cite{Woyn} regarding a far-reaching
simplification in the Lieb and Wu Bethe-Ansatz solution of
the Hubbard model\cite{Lieb} in the $U \rightarrow \infty$ limit. This
Woynarovich-Ogata-Shiba work just amounts to the realization
that in the large $U$ limit the structure of the Bethe-Ansatz
solution coincides with a squeezed space construction. For 
simplicity, assume a  thermodynamical potential $\mu > 0$
such that no doubly occupied sites occur. For $U$ tending to 
infinity, the ground state wavefunction $\psi$ of a Hubbard
chain of length $L$ occupied by $N$ electrons (with $N < L$)
factorizes into a simple product of spin- and charge wavefunctions,
\begin{widetext}
\begin{equation}
\psi(x_1,\ldots ,x_N \; ; \; y_1, \ldots , y_{M}) =  \psi_{SF} (x_1, \ldots, x_N)
\quad \psi_{Heis.} (y_1, \ldots, y_{M}) \; .
\label{Deb}
\end{equation}
\end{widetext}
The charge part $\psi_{SF}$ represents the wave function of
 non-interacting spinless
fermions where the coordinates $x_i$ refer to the positions of the 
$N$ singly occupied sites. The spin wavefunction  $\psi_{Heis.}$ is 
identical to the wave function of a chain of Heisenberg spins interacting 
via an  antiferromagnetic nearest neighbor exchange, and the
coordinates $y_j$, $j=1, \ldots, M$ refer to the $M$ positions occupied
by the up spins in the Heisenberg chain. The surprise is that the coordinates 
$y_j$ do not refer to the original Hubbard chain with length $L$,
 but instead to a new space: a chain of length $N$ constructed from the 
 sites at coordinates $x_{1},  x_{2}, ..., x_{N}$ given by  
the positions of the charges (singly occupied sites) in a configuration with 
amplitude $\psi_{SF}$.  One immediately notices that it is identical to the 
squeezed space mapping for the Haldane spin chains discussed in the previous 
section, associating the $M_s = 0$ states of the spin chain with the
holes and the $M_s = \pm 1$ states with the singly occupied sites carrying
electron spin up $(+)$ or down $(-)$. In fact, as already pointed out by Batista and
Ortiz\cite{Batista}, one can interpret the spin chain as just a bosonic
$t-J_z$ mode, i.e. lowering the $SU(2)$ symmetry of the Hubbard model
to Ising, dismissing the Jordan-Wigner strings making up the difference
between spinless fermions and hard-core bosons, and last but not least
adding an external Josephson field forcing the holes ($M_s = 0$, in the spin language) 
to condense giving a true Bose condensate.

Since the geometrical mapping is the same, a `string' operator equivalent
to that of den Nijs and Rommelse can be constructed for the Luttinger
liquid. In order to measure the spin correlations in squeezed space starting
from unsqueezed space one should construct an operator which removes the
sublattice parity flips. Define the staggered magnetization in unsqueezed 
space as,  
\begin{equation}
\vec{M} (x) =  (-1)^{x} \vec{S} (x) \; .
\label{stagmag}
\end{equation}
Compared to the corresponding quantity in squeezed space, these acquire an 
additional fluctuation  due to the motions of the sublattice parity flips.
Since these flips are attached to the holes, they can be `multiplied out'
by attaching a `charge-string',
\begin{equation}
(M')^z(x) = M^z (x) (-1)^{\sum_{j=- \infty} ^{x-1}(1- n_{tot}(j))} \; ,
\label{sqstagmag}
\end{equation}
where $1 -n_{tot}(j)$ is the number of holes on site $j$
 and the charge operator 
$n_{tot}(j) = n_{\uparrow}(j) + n_{\downarrow}(j)$ taking the values 0,
1 and 2 for an empty-, singly- and doubly occupied site, respectively.
$(M')^z$ is representative for  the `true' staggered magnetization living
in squeezed space. The action of the charge string $\Pi_j (-1)^{1 - n_{tot} (j)}$
is to add a $-1$ staggering factor only when the site $j$ is singly occupied,
thereby reconstructing the bipartiteness in squeezed space. It follows that
the analogue of the den Nijs topological operator becomes,
\begin{equation}
\begin{array}{rcl}
\Ostr (x) &=& 
\langle ~( M')^z(x)~~  ( M')^z(0)  ~\rangle \\
&=&
\langle M^{z} (x)
(-1)^{ \sum_{j=0}^{x-1} ( 1 - n_{tot}(j) ) } M^{z} (0)  \rangle \\
&=& - \langle S^z(x) (-1)^{ \sum_{j=1}^{x-1}  n_{tot}(j)  }   S^z(0) \rangle \; .
\label{otop}
\end{array}
\label{eq:stringcorrdefn}
\end{equation}

The focus of the remainder of the paper is on the analysis of this correlator.
To the
best of our knowledge, correlators of this form have only been considered
before in the context of stripe fluids in 
2+1D\cite{Zaanenphilmag,ZaanenSaarloos}. 
String correlators have been constructed before in the one dimensional
context\cite{Resta,Talstra} 
but these are of a different nature, devised to detect `hidden
order' of an entirely different type. 

On this level of generality it might appear that the hidden
order of the Haldane chain duplicates that of the Luttinger
liquid. However, dynamics matters and in this regard the Luttinger
liquid is quite different. Instead of genuine disorder in the
`charge' sector and the true long range order in the `spin' sector
of the spin chain, both charge and spin are critical in the Luttinger
liquid and this makes matters more delicate.

We learned in the previous section that in order to 
measure the hidden order one should compare 
the conventional two point spin correlator $\langle \vec{M} (r) \vec{M} (0)
\rangle$ with  the string correlator defined in Eq. (\ref{eq:stringcorrdefn}). 
Let us compute these correlators
explicitly in the large $U$ limit. In the calculation, the string correlator 
turns out to be a simplified version of the two point correlator. The latter
was already computed by Parola and Sorella\cite{Parola} 
starting from the squeezed space
perspective. Let us retrace their derivation to find out where the
simplifications occur. 

Start with the observation that a Heisenberg spin antiferromagnet
 is realized in squeezed space.
This implies that the squeezed space spin-spin correlator
has the well-known asymptotic form, 
\begin{equation}
\begin{array}{rcl}
O_{Heis.}(j) & \equiv & \langle S^z (j) S^z (0) \rangle \\ 
& \rightarrow &  (-1)^j
\Gamma \; \; \frac{\ln^{1/2}(j)}{j} \\
& \equiv & (-1)^j O_{stag} (j)
\label{Heiscor}
\end{array}
\end{equation}
where $\Gamma$ is a constant\cite{Singh}, while $j$ labels the sites
in squeezed space. 

The charge dynamics are governed by an effective
system of non-interacting spinless fermions. Define their number 
operators as $n (l)$ where $l$ refers to sites in full space. Define
the following correlation function, to be evaluated relative to the
spinless fermion ground state,
\begin{equation}
\label{caked}
P^x_{SF} (j) = \langle n(0) n(x) \delta \left( 
\sum_{l=0}^{x} n(l) -j \right)\rangle _{SF} \; .
\label{parsorsq}
\end{equation}
By definition this measures the probability of finding $j$ spinless
fermions in the interval $\left[0,x\right]$, given one fermion located at site $0$ and one 
at site $x$. Parola and Sorella\cite{Parola} show that the
exact relation between Eq. (\ref{Heiscor}) and the two point correlator in
full space is,
\begin{equation}
\begin{array}{rcl}
\langle S^z(x)S^z(0) \rangle & = & \displaystyle \sum_{j=2}^{x+1}P^x_{SF} (j) O_{Heis.} (j-1)
 \vspace{1mm} \\
& = & \displaystyle \sum_{j=2}^{x+1}P^x_{SF} (j) (-1)^{j-1} O_{stag} (j-1) \vspace{1mm} \\
& \rightarrow & -  \displaystyle \sum_{j=2}^{x+1}P^x_{SF} (j) (-1)^{j} O_{stag} (j) \; .
\end{array}
\label{pasots}
\end{equation}
Let us now consider instead the string correlator,
\begin{equation}
\begin{array}{rcl}
\Ostr (x) & = &
- \langle S^z(0) (-1)^{\sum_{j=1}^{x-1}  n(j)  } S^z(x) \rangle 
 \vspace{1mm} \\
 & = & - \displaystyle \sum_{j=2}^{x+1} P_{SF}^x (j)
(-1)^{j-2} O_{Heis.} (j-1)
   \vspace{1mm} \\
& \rightarrow & \displaystyle \sum_{j=2}^{x+1} P_{SF}^x (j)
O_{stag} (j) \; .
\end{array}
\label{pasotop}
\end{equation}
The difference between the two point correlator and the string
correlator looks at first sight to be rather unremarkable. The staggering
factor $(-1)^j$ associated with the sign of staggered spin in squeezed
space (Eq. \ref{Heiscor}) survives for the two point correlator, but
it is canceled for the topological correlator because $(-1)^{j-2}
\times (-1)^{j-1} = (-1)^{2j - 3} = -1$. However, this factor
is quite important because it is picked up by the charge sector 
due to the $\delta$ function appearing in the definition of
$P_{SF}$ (Eq. \ref{parsorsq}).

In Eqn.'s (\ref{pasots},\ref{pasotop}) spin and charge are still `entangled'
due to the common dependence on $j$. However, it can be demonstrated
that asymptotically this sum factorizes. It can be proven\cite{Parola} that
the sum $\sum_{j=2}^{x+1} P_{SF}^x(j) (-1)^j f(j)$ with  $f(j)$ bounded and
satisfying
\begin{equation}
\left| \frac{f(j) - f(j')}{j-j'} \right| \leq 2 \Gamma 
\frac{\ln^{1/2} (x)}{x^2}
\label{proet}
\end{equation}
differs from the sum
\begin{equation}
\left[ \sum_{j=2}^{x+1} P_{SF}^x(j) (-1)^j \right] f(\langle r \rangle_x)
\label{proet1}
\end{equation}
where ($\rho_{tot} = N /L$ is the fermion density) 
\begin{equation}
\begin{array}{rcl}
\langle r \rangle_x &=& \frac{1}{\langle n(0) n(x)\rangle _{SF}} 
\sum_{j=2}^{x+1} j ~ P^x_{SF} (j)  \\
 & = & x \rho_{tot} +1 \rightarrow x \rho_{tot}
\end{array}
\label{raver}
\end{equation}
by terms vanishing faster than  $\frac{\log^{3/2}(x) }{x^2}$. 
The condition Eq. (\ref{proet}) is satisfied by the squeezed space staggered
magnetization  $f(j) \sim O_{stag} (j)$ and since the above result does not
depend on the presence of the staggering factor $(-1)^j$ it applies equally
well to the two point spin correlator and the string correlator.

Given this factorization property, let us first consider the 
string correlator,
\begin{equation}
\begin{array}{rcl}
\Ostr (x) & = & \displaystyle \sum_{j=2}^{x+1} P_{SF}^x (j)
O_{stag} (j) 
 \\
& = & \left[ \displaystyle \sum_{j=2}^{x+1} P_{SF}^x (j)
 \right] O_{stag} (x \rho_{tot} )  
\\ & & \quad
+ {\cal O}\left(        
\frac{\ln^{3/2} (x)}{x^2}
\right) 
\end{array}
\label{otopx1}
\end{equation}
It is easy to demonstrate that the sum over the $P_{SF}$ is just the
density-density correlator of the non-interacting spinless fermion
system 
\begin{equation}
\begin{array}{rcl}
\sum_{j=1} ^{x+1} P_{SF}^x (j) &  = & \langle n(0) n(x)\rangle _{SF} 
 \\
& = & \rho_{tot}^2 - \frac{1}{2} \left(  \frac{1-\cos (2k_F x)}{\pi x^2}\right),
\label{sfdens}
\end{array}
\end{equation}
with $k_F = \pi \rho_{tot}$. We arrive at the simple exact result,
\begin{equation}
\begin{array}{rcl}
\Ostr (x) & = &  \langle n (x) n (0) \rangle_{SF}
   \frac{\Gamma}{ \rho_{tot}x} \ln^{1/2} (\rho_{tot} x)
 + {\cal O}\left(        
\frac{\ln^{3/2} (x)}{x^2}
\right)  \\
&= &  \Gamma  \frac{\rho_{tot}}{x} \ln^{1/2} (\rho_{tot} x)
 + {\cal O}\left(        
\frac{\ln^{3/2} (x)}{x^2}
\right).
\end{array}
\label{toplUres}
\end{equation}

This confirms the intuition based on the squeezed space picture. 
The topological correlator just measures the spin correlations
in squeezed space which are identical to those of a Heisenberg
spin chain, Eq. (\ref{Heiscor}). At short distances this is not
quite true, but it becomes precise at large distances due to the
asymptotic factorization property Eq. (\ref{otopx1}). Of course,
$\Ostr$ measures in units of length of the full space and because
in squeezed space sites have been removed the unit of length is
uniformly dilated $x \rightarrow \rho_{tot} x$. By the same token,
the amplitude factor reflects the fact that there are only $\rho_{tot}$
spins per site present in full space.

The calculation of the two point spin correlator is less easy. Using
again the factorization property,
\begin{widetext}
\begin{equation}
\begin{array}{rcl}
\langle S^z(x)S^z(0)\rangle & = & - \displaystyle \sum_{j=2}^{x+1}P^x_{SF} (j) \; (-1)^{j} \, f(j) 
 \\
& = & - \left[ \displaystyle \sum_{j=2}^{x+1}P^x_{SF} (j) (-1)^{j}\right] O_{stag}(\langle r
\rangle_x ) + 
{\cal O} \left(\frac{\ln^{3/2} (x)}{x^2}\right)
 \\
 & = &  \displaystyle - D_{nn,SF} (x) \; \Gamma \frac{\ln^{1/2} (\rho_{tot} x)}{\rho_{tot} x}+ 
{\cal O} \left(\frac{\ln^{3/2} (x)}{x^2}\right).
\label{SSint1}
\end{array}
\end{equation}
\end{widetext}
Due to the staggering factor, the `charge function' $D_{nn,SF} (x)$ is now more
interesting,
\begin{equation}
\begin{array}{rcl}
D_{nn,SF} (x) & = & \displaystyle \sum_{j=2}^{x+1}P^x_{SF} (j)~ (-1)^j \\
& = & \displaystyle \sum_{j=2}^{x+1} \langle n(0) n(x)
 \delta \left( \sum_{l=0}^x n(l) -j \right) \rangle_{SF}~  (-1)^j
 \\
&=&
\langle n(0) (-1)^{ \sum_{l=0}^x n(l)}   n(x) \rangle_{SF}
\end{array}
\label{DnnSFdef}
\end{equation}

The spin correlations are modulated by a function reflecting the uncertainty
in the number of sublattice parity flips which can be expressed in terms of
expectation values of charge string operators. For spinless fermions the
following exact identity holds for the number operator,
\begin{equation}
n (j) = \frac{1}{2} \left[ 1 - (-1)^{n (j)} \right]
\label{spferid}
\end{equation}
which implies,
\begin{equation}
D_{nn,SF} = \frac{1}{4} \left[ D_{SF} ( x - 2 ) + D_{SF} ( x ) - 2 D_{SF} ( x - 1) \right]
\label{DnntoD}
\end{equation}
demonstrating that this function is the second lattice derivative of
the charge-string correlator,
\begin{equation}
D_{SF} (x) \equiv \langle   (-1)^{ \sum_{l=0}^x n(l)} \rangle_{SF} \; .
\label{Ddef}
\end{equation}

Even for free spinless fermions this function has not been derived 
in closed analytic form. However, it can be easily evaluated numerically
and we show in the appendix that it is very accurately approximated by,
\begin{equation}
\langle \, (-1)^{\sum_{j=1}^{x-1} n_(j) } \,
\rangle_{SF}  = 
 \frac{A^2 \sqrt{2}}{\sqrt{\sin( \pi \rho_{tot} )}}
 \frac{ \cos( \pi \rho_{tot} (x-1) ) }{\sqrt{x-1}} \; ,
\label{chstriSF}
\end{equation}
where $A$ is a constant evaluated to be $A=0.6450002448$\cite{Parola}. 
Using Eq. (\ref{DnntoD}) it follows immediately that,
\begin{equation}
\begin{array}{rcl}
D_{nn,SF} ( x )  & =  & \displaystyle
\langle \, n (x) \, (-1)^{ \sum_{j=1}^{x-1}  n (j) } \, 
n (0) \, \rangle_{SF} 
 \\
& = & \displaystyle \frac{A^2 (\cos(\pi \rho_{SF})-1 )}{ \sqrt{2 \sin(\pi \rho_{SF})}}
\frac{\cos(\pi \rho_{SF} x)}{ \sqrt{  x }}
 \\
&  \rightarrow &  \displaystyle { {A^2} \over {\sqrt{2} }} { {\cos( 2k_F x )} \over 
{ x^{K_c} } } \,
\end{array}
\label{chstrnSF}
\end{equation}
where, as before, $2k_F = \pi \rho_{tot}$ and introducing the charge
stiffness $K_c$ which takes the value $1/2$ in a free spinless fermion
system. This is the desired result, and combining it with Eq. 
(\ref{SSint1}) we arrive at the asymptotically exact result for the
two point spin correlator in the large $U$ limit,
\begin{equation} 
\begin{array}{rcl}
\langle S^z(x) S^z(0)\rangle & = & \displaystyle A^2 \, \sqrt{2} \, \Gamma 
\frac{\cos (2k_F x)}{\rho \, x^{ 1 + K_c}}
\ln^{1/2}
(x/2) 
\\ & & \quad 
\displaystyle + {\cal O} \left(\frac{\ln^{3/2} (x)}{x^2}
\right) \; .
\end{array}
\label{SSlargeU}
\end{equation}

This calculation demonstrates quite explicitly why the spin correlations
in this Luttinger liquid are sensitive to the charge fluctuations. The
latter enter via the uncertainty in the location of the sublattice
parity flips which is expressed via the function $D_{nn}$ or equally
the more fundamental function $D$. Due to the factorization property
Eq. (\ref{SSint1}) it enters in a multiplicative fashion. The 
string correlator is constructed to be insensitive to the
sublattice parity fluctuation and it follows that,
\begin{equation}
\langle S^z(x) S^z(0)\rangle \; \propto \; { 1 \over { x^{K_c}} } \;
\langle S^z(x) (-1)^{ \sum_{j=1}^{x-1}  n_{tot}(j)  }   S^z(0) \rangle \; .
\label{SStotop}
\end{equation}
This is in close analogy with Eq. (\ref{wilsonl}) for the Haldane chain. 
The difference is that in the spin chain the string correlator
is decaying exponentially slower than the two point correlator
while in the large $U$ Luttinger
liquid the difference is only algebraic. This has an obvious reason.
In the spin chain, the `charge' sector is truly disordered 
(Bose condensed), such
that the `charge-charge' correlations decay exponentially and this
will obviously also cause an exponential decay of the charge string
correlator $D$. The charge sector in the Luttinger liquid is critical,
 exhibiting algebraic correlations. As have demonstrated explicitly
above, this also renders  $D$ to be algebraic. We argued in
section \ref{sec:gauge} that the exponential difference found in the spin chain signals the
emergence of an Ising  gauge symmetry: the charge string just corresponds
with the Wilson loop of the gauge theory. By the same token, the algebraic 
difference in the Luttinger liquid means that the Ising gauge symmetry
is not quite realized. However, power laws indicate criticality and this
is in turn associated with  a second order phase transition. Thus we are
considering a correlator which measures directly the gauge fields; its
power law characteristic indicates that the gauge symmetry itself is 
involved, and the logical consequence is: 
{\em the Luttinger liquid is located at the continuous
phase transition where local Ising symmetry emerges.}

This sounds odd at first sight. However, one should realize that this
Ising gauge symmetry is just dual to the superfluid phase order in the
charge sector. Although in 1+1D  true long range superfluid order cannot 
exist, the Luttinger liquid can be viewed as an entity which is at the
same time an algebraic superfluid and an algebraic charge density wave.
In principle, when one applies an external Josephson field acting on the
charge sector alone it will directly turn into a true superfluid. In
this superfluid the number correlations are short ranged and this implies
that the charge-string will decay exponentially. 

A caveat is that this
Josephson field has to be applied in such a way that the spin system is
unaffected. 
For instance, applying a standard Josephson field acting say on the
singlet channel $\sim \Psi_{\uparrow} \Psi_{\downarrow}$ has the 
automatic effect that a spin gap opens and one can continue
adiabatically to the strong singlet pairing limit. At long distances,
all that remains is doubly occupied sites and holes and it is no longer
possible to construct squeezed space. It is `eaten' by the spin gap. 
However, at least in principle one can construct a `charge only'
Josephson field. Consider the large $U$ limit.
 The Bethe-ansatz wavefunction
demonstrates that the ground state in the decoupled charge sector is
in one-to-one correspondence to that of  a free spinless fermion Hamiltonian.
One can simply add to this Hamiltonian a Josephson field acting directly
on the spinless fermions $\sim H_J \sum_{<ij>} c^{\dagger}_i c^{\dagger}_j$
and for any finite strength of $H_J$ the charge ground state will 
correspond with a BCS superconducting state. By construction, this field
will leave the squeezed space structure and the spin sector unaffected.
The ramification is that the quantization of number density is truly 
destroyed and since holes continue to be bound to the sublattice parity,
the disorder in the number sector becomes the same as $Z_2$ gauge degeneracy
in the spin sector. This is the same type of construction as  suggested
by Batista and Ortiz\cite{Batista} in their identification 
of the Haldane spin chain with a {\em superfluid} $t-J_z$ model.

\section{Squeezed space and non-interacting electrons.}
\label{sec:NonInteracting}

The existence of squeezed space is remarkable, and intuitively
one might think that one needs highly intricate dynamics associated
with strong electron-electron interactions in order for squeezed
space to have a chance to emerge. The evidence for its existence 
presented so far is
entirely based on very special strongly interacting cases (the
Haldane spin chain, the large $U$ Hubbard model) which can be solved
exactly for more or less accidental reasons. However,
in the previous paragraphs we have constructed and tested a measuring
device which can unambiguously detect squeezed space also in cases
where simple exact wavefunctions are not available. Alternatively,
it can be detected even in cases
where one knows the wavefunction but where the squeezed space structure
is deeply buried because the coordinates are not of the right kind.
Our measuring recipe is straightforward: compute the string
correlator Eq.(\ref{otop}) and find out if behaves like the pure
spin chain, or whatever `matter' system one expects to populate
squeeze space.

The simplest possible example is the non-interacting, spinful 
electron system. As we will demonstrate using only a few lines
of algebra, it
survives the test! We interpret this as a remarkable feat
of the fermion minus signs. Squeezed space refers eventually 
to a bosonic representation of the fermion problem, and apparently
the  minus sign structure in terms of the fermion representation
is of sufficient complexity to make possible an entity as organized
as squeezed space in the boson language.

The proof is as follows. For a system of $S=1/2$ fermions we can use 
the following operator relations,
\begin{equation}
\begin{array}{rcl}
S^z(y) & = & \frac{1}{2} (n_\uparrow(y) - n_\downarrow(y) )
 \\
n_{tot}(y) & = & n_\uparrow(y) + n_\downarrow(y) \; .
\label{sndef}
\end{array}
\end{equation}
The string correlator can be written as,
\begin{widetext}
\begin{equation}
\begin{array}{rcl}
\Ostr(x) & = &
- \langle \, S^z (x) \, 
(-1)^{\sum_{j=1}^{x-1}  n_{tot}(j) } \, S^z(0) \, \rangle  \\ 
& = &
  - \frac{1}{4}\langle \,  n_{\uparrow } (x) \,
(-1)^{\sum_{j=1}^{x-1} n_{\uparrow}(j)} \,
 n_{\uparrow} (0)\, 
 \rangle
\langle \, (-1)^{\sum_{j=1}^{x-1}  n_{\downarrow}(j) } \, \rangle
% \\ & &
 -  \frac{1}{4}
 \langle \, n_\downarrow (x) \,
  (-1)^{\sum_{j=1}^{x-1}
 n_{\downarrow}(j)}\,
n_\downarrow (0)\,
  \rangle
\langle \, (-1)^{\sum_{j=1}^{x-1}  n_{\uparrow}(j) } \, \rangle
 \\
& & + \frac{1}{4} \langle \, n_\uparrow (x)  \,
(-1)^{\sum_{j=1}^{x-1} n_{\uparrow} (j)} 
\,  \rangle
\langle \, n_\downarrow(0) \, 
(-1)^{\sum_{j=1}^{x-1}  n_{\downarrow}(j) } \, \rangle
% \\ & & 
+ \frac{1}{4}\langle \, n_\downarrow (x) \,
(-1)^{\sum_{j=1}^{x-1} n_{\downarrow} (j)}
\,  \rangle
\langle  \,
n_\uparrow(0) \,
(-1)^{\sum_{j=1}^{x-1}  n_{\uparrow}(j) } \, \rangle \; .
\end{array}
\label{orem}
\end{equation}
\end{widetext}

In the non-interacting limit, the spin up and spin down electrons
behave as two independent species of free spinless fermions. Since
the expectation value of any operator involving only either up- or down
spin creation and annihilation operators is the same, Eq. (\ref{orem})
simplifies to,
\begin{widetext}
\begin{equation}
\begin{array}{rcl}
\Ostr & = &
- \langle \, S^z (x) \, 
(-1)^{\sum_{j=1}^{x-1}  n_{tot}(j) } \, S^z(0) \, \rangle  =
  \\
 & & - \frac{1}{2}\langle \,  n_{SF } (x) \,
 (-1)^{\sum_{j=1}^{x-1} n_{SF}(j)} \,
 n_{SF} (0)\,
 \rangle
\langle \, (-1)^{\sum_{j=1}^{x-1}  n_{SF}(j) } \, \rangle
 \\
& & + \frac{1}{2} \langle \, n_{SF} (x)  \,
(-1)^{\sum_{j=1}^{x-1} n_{SF} (j)} 
\,  \rangle
\langle \, n_{SF}(0) \, 
(-1)^{\sum_{j=1}^{x-1}  n_{SF}(j) } \, \rangle \; .
 \\
\end{array}
\end{equation}
\end{widetext}
where the operators now refer to spinless fermions. We recognize 
in this expression the $D_{SF}$ and the $D_{nn,SF}$ we already
encountered in section \ref{sec:luttinger} (Eqn.'s \ref{DnnSFdef},\ref{Ddef}).
In addition we also need,
\begin{equation}
\begin{array}{rcl}
D_{n, SF} & = & \langle \, n_{SF} (x)  \,
(-1)^{\sum_{j=1}^{x-1} n_{SF} (j)} 
\,  \rangle
 \\
& = & { 1 \over 2} \left( D_{SF} ( x - 2 ) - D_{SF} ( x - 1 ) \right)
\end{array}
\label{DnSFdef}
\end{equation}

being the first lattice derivative of $D$, using once again 
the operator identity Eq. (\ref{spferid}). The topological
correlator can therefore be expressed entirely in terms
of the `fundamental' string operator $D _{SF} (x) \sim \langle
\Pi (-1)^{n_{SF}} \rangle$ as,
\begin{equation}
 \Ostr (x) =  \frac{1}{8} \left[  D_{SF} (x-2) D_{SF} (x) -D_{SF} (x-1)^2
 \right]
\label{otfree1}
\end{equation}

The function $D_{SF} (x)$ was already encountered (Eq. \ref{chstriSF},
see also the appendix) and using this result,
\begin{equation}
\begin{array}{rcl}
\Ostr (x)  &=& \frac{A^4 \sin(\pi \rho_{SF})  }{4 x }
 \\ 
& = & \frac{A^4 \sin(k_F)  }{4 x }
 \\
&=& \frac{A^4 \sin(\pi \rho_{tot}/2)  }{4 x },
\end{array}
\label{freeltop}
\end{equation}
where 
$\rho_{SF} = {\rho_{tot}}/{2} = ({\rho_{\uparrow} +\rho_{\downarrow} })/{2}$.
Note that $2 k_F = \pi \rho_{tot} = \pi \frac{2 N_{SF}}{V}$ and 
so $k_F = \frac{\pi N_{SF}}{V} = \pi \rho_{SF}$. We also calculated
the string correlator numerically using the method explained in
the appendix. 

\begin{figure}[tbh]
\includegraphics[angle=270,width=\figurewidth]{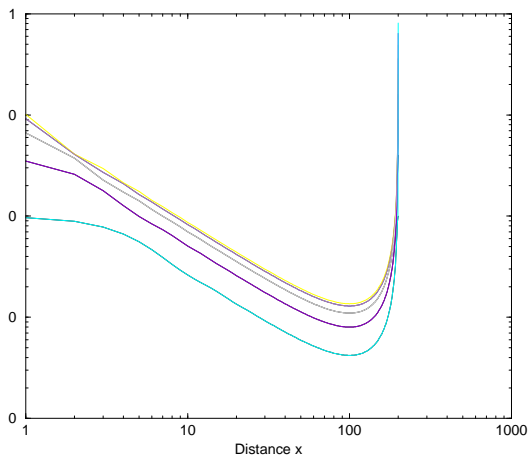}
\caption{The function 
$\Ostr = -\langle S^z(x) (-1)^{\sum_{j=1}^{x-1} n_{tot}(j)} S^z(0) \rangle$ 
as computed numerically for the free spinful fermion
gas at densities 
$\rho_{tot} =0.2$, $\rho_{tot} =0.6$ and  $\rho_{tot} =1$,
shown in a log-log plot. The algebraic decay implied by Eq. (\ref{freeltop})
is indicated by the straight line.}
\label{allesbijeen}
\end{figure}

In figure \ref{Deind20} we show the numerical result for $\Ostr(x)$
for a density $\rho_{tot} = 2 {N_{SF}}/{V}  = 0.2$ and $V=200$ which
is in excellent agreement  with the analytic expression Eq. (\ref{freeltop}).
In Figure \ref{allesbijeen} we show the numerical results for various
densities on a  log-log plot highlighting the algebraic decay with an
exponent $K_s = 1$. 

\begin{figure}[tbh]
\includegraphics[angle=270,width=\figurewidth]{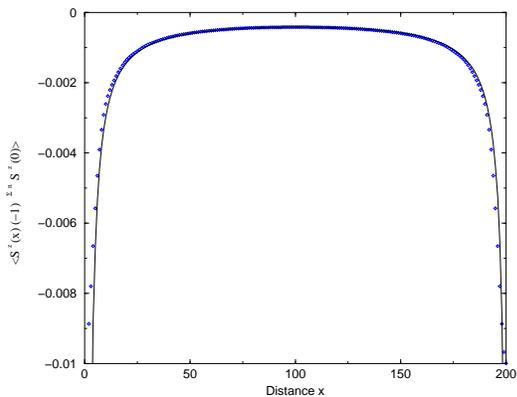}
\caption{The function 
$\Ostr= -\langle S^z(x) (-1)^{\sum_{j=1}^{x-1} n_{tot}(j)} S^z(0) \rangle$ 
for $U=0$ calculated numerically using the
algorithm discussed in appendix A.
 Here $\rho_{tot} = 2 {N_{SF}}/{V} = 0.2$ and $V=200$.
The drawn line is the analytic solution Eq. (\ref{freeltop}).}
\label{Deind20}
\end{figure}

It is obvious where this exponent, equal to unity, is coming from in
the calculation. From Eq. (\ref{otfree1}) it follows that
$\Ostr \sim 1 / (x^{K_{c,SF}})^2$ where the spinless fermion
exponent $ K_{c,SF} = 1/2$. This looks at first sight rather
unspectacular but one has to realize that the two point spin
correlator of the free-fermion gas decays faster, $\langle
S S \rangle \sim 1 / x^2$ and the topological correlator therefore
uncovers a more orderly behavior. Furthermore, the only symmetry 
reason to expect such an exponent to be equal to unity is
the protection coming from $SU(2)$ (spin) symmetry. Can we be
certain that this result proves that even in the non-interacting 
limit a Heisenberg chain is lying within squeezed space? The
above computation is not very explicit in this regard and the
persuasive evidence is still to come: bosonization, and especially
the numerical results presented in section \ref{sec:numerical} showing that the asymptotic
behavior of the string correlator is independent of $U$ and density.  

\section{Squeezed spaces and bosonization.}
\label{sec:bosonization}

Arriving at this point, we are facing evidence that the squeezed
space is actually not at all special to the large $U$ limit. It
could well be ubiquitous in one dimensional electron systems. How
does bosonization fit in? After all, during the last thirty years
overwhelming evidence accumulated for bosonization to be the
correct theory in the scaling limit. Squeezed space is of course 
fundamental; it is among others a precise description of the
meaning of spin-charge separation. How could bosonization ever be
correct if it would not somehow incorporate the squeezed space
structure? In section \ref{subsec:vertex} we make the case that the peculiarities
in the structure of the theory, originating in the core of
the bosonization `mechanism' (i.e., the Mandelstam construction for 
the fermion operators), are just coding for squeezed space. Again,
the string operator is the working horse. By just tracking the
fate of the string- and two point spin correlators in the 
bosonization framework, it becomes evident that it is in one-to-one
correspondence with the strong coupling limit. This observation is
further amplified in Appendix B where we discuss an intuitive 
argument by Schulz which turns out to subtly misleading.  To fix 
conventions, let us start out collecting some standard expressions.

\subsection{The bosonization dictionary.}
\label{subsec:dictionary}

To fix conventions let us collect here the various standard
bosonization expressions we need later\cite{VoitSchulz,Stone}. 
At the Tomonaga-Luttinger
fixed point the dynamics is described in terms of gaussian scalar fields 
$\varphi_s$ and $\varphi_c$ for spin and charge, respectively. Introducing
conjugate momenta $\Pi_{s,c}$ the Hamiltonian is,
\begin{equation}
\begin{array}{rcl}
H_{TL} &=& \sum_{\mu=c,s} \frac{v_\mu}{2}
\int dx
 \left[
K_\mu \Pi^2_\mu + \frac{1}{K_\mu} (\partial_x \varphi_\mu)^2
 \right],
\end{array}
\label{TLham}
\end{equation}
where $K_s$ ($v_s$) and $K_c$ ($v_c$) are the spin and charge stiffness 
(velocity) respectively. For
globally $SU(2)$ symmetric spin systems $K_s = 1$ and  $K_c$ 
is depending on the microscopy, but generally $0 \leq K_c < 1$ for
repulsive interactions.

Electron operators can be re-expressed in terms of these
bosonic fields via the Mandelstam construction. Starting from 
the spinful Dirac Hamiltonian describing the linearized 
electron-kinetic energy,
\begin{equation}
H_0 = - i v_F \sum_{\sigma} \int dx \left[ \psi^{\dagger}_{\sigma} ( x )
\partial_x \psi_{\sigma} ( x ) - \bar{\psi}^{\dagger}_{\sigma} ( x ) 
\partial_x \bar{\psi}_{\sigma} ( x ) \right]
\label{dirac}
\end{equation}

The field operators of the left- ($\bar{\psi}_{\sigma}$) and right
($\psi_{\sigma}$) moving fermions are expressed in terms
of the bose fields as,
\begin{equation}
\begin{array}{rcl}
\psi_{\sigma} (x)  & = & { {\eta_{\sigma} } \over {\sqrt{2 \pi} } }
\; e^{ i \sqrt{\pi} \left( \varphi (x) - \int_{-\infty}^x dy \Pi ( y ) \right) }
 \\
 \bar{\psi}_{\sigma} ( x ) & = &  { {\bar{\eta}_{\sigma} } \over {\sqrt{2 \pi} } }
\; e^{ - i \sqrt{\pi} \left( \varphi (x) + \int_{-\infty}^x dy \Pi ( y ) \right) } \; ,
\end{array}
\label{mandelstam}
\end{equation}
where $\eta_{\sigma}, \bar{\eta}_{\sigma}$ are the Klein factors keeping
track of the fermion anti-commutation relations.

Starting from the normal ordered charge density
the total charge density can be written as,
\begin{equation}
\begin{array}{rcl}
n_{tot} (x) & = & :n_{\uparrow} (x) + n_{\downarrow} (x): \\
& \simeq & 
\sqrt{\frac{2}{\pi}} \frac{\partial \varphi_c}{\partial x} +
{\cal O}_{CDW}(x)
+{\cal O}^\dagger_{CDW}(x) \; ,
\end{array}
\label{ntotbos}
\end{equation}
where $\partial_x \varphi_c$ represents uniform components of the charge
density, while the various finite momentum components
are lumped together into $O_{CDW}$. The dominant contributions come from momenta
$q = 2k_F$ and  $4k_F$,
\begin{equation}
\begin{array}{rcl}
{\cal O}_{CDW}(x) & = & {\cal O}_{2k_F}(x) + {\cal O}_{4k_F}(x) \\
 {\cal O}_{2k_F}(x) & = & 
\frac{1}{\pi} e^{-2ik_F x} e^{i\sqrt{2\pi} \varphi_c(x)}
\cos\left[ \sqrt{2\pi}  \varphi_s(x) \right]
 \\
{\cal O}_{4k_F}(x) & = & e^{-4ik_Fx} \frac{1}{2\pi^2} e^{i \sqrt{8 \pi}
\varphi_c(x) } \; .
\end{array}
\label{OCDWdef}
\end{equation}
Similarly, the spin operator $S^z (x)$ becomes,
\begin{equation}
\begin{array}{rcl}
S^z (x) &=&\frac{ 
: n_\uparrow(x) -
 n_\downarrow(x) :
}{2} \\
&=& 
 \sqrt{\frac{1}{2\pi}} \partial_x \varphi_s(x) +{\cal O}_{SDW,z}(x)
+{\cal O}^\dagger_{SDW,z}(x) \; ,
\end{array}
\label{Spindef}
\end{equation}
where $\partial_x \varphi_s$ refers to the uniform (ferromagnetic)
component while the finite wavevectors are dominated by the
$q= 2k_F$ component,
\begin{equation}
\begin{array}{rcl}
{\cal O}^\dagger_{SDW,z}(x)& \simeq &  {\cal O}_{Sz, 2k_F} (x) 
 \\
&=& \frac{i}{2 \pi} e^{-2ik_F x} e^{i\sqrt{2 \pi} \varphi_c(x) }
 \sin \left[ \sqrt{2 \pi} \varphi_s(x) \right] \; .
\end{array}
\label{OSDWdef}
\end{equation}

In addition we need the usual rules for constructing the propagators
of (vertex) operators in a free field theory like Eq. (\ref{TLham}), 
\begin{equation}
\begin{array}{rcl}
\langle \partial_x \varphi_{\mu} (x)  \partial_x \varphi_{\nu} (0) \rangle
& = & - \delta_{\mu, \nu} { { K_c} \over {2\pi} } { 1 \over {x^2} }
 \\
\langle e^{i n \sqrt{ 2 \pi} \left[  \varphi_{\mu} (x) -  \varphi_{\mu} (0) \right] }
\rangle 
& = & { 1 \over { x^{n^2 K_{\mu} } } } \; .
\end{array}
\label{propgaus}
\end{equation}

\subsection{Vertex operators and squeezed space.}
\label{subsec:vertex}

It is a peculiarity of bosonization  that the
charge field enters the spin sector in the form of a vertex
operator $\sim e^{i\varphi_c}$, see Eq.(\ref{OSDWdef}). This can
be traced back to the Mandelstam construction for the
fermion field operators, Eq. (\ref{mandelstam}),
indicating that the fermions are dual to the fields
$\varphi$: the fermions have to do with solitons or kinks in the bose
fields. 

Let us observe the workings of bosonization from the viewpoint
offered by the strong coupling limit discussed
in section \ref{sec:gauge}. We found that the charge-string correlator
$D(x)$ is the most fundamental quantity keeping track of
the fluctuations in the sublattice parity. Let us see
what bosonization has to say about this correlator.

This function becomes in  the continuum,
\begin{equation}
\begin{array}{rcl}
D (x) & \equiv &  \langle (-1)^{\sum_{j=0}^{x} n_{tot}(j)} \rangle
 \\
& = & \langle  \cos \left[ \pi \sum_{j=0}^x n_{tot} (j) \right] \rangle
 \\
& \rightarrow & \langle         
\cos \left[ \pi \int_{0}^{x} dy ~ n_{tot}(y) \right] \rangle.
\label{contD}
\end{array}
\end{equation}

The theory is constructed to represent the scaling limit and
therefore we should focus on the leading singularities. 
According to Eq.(\ref{ntotbos}), the total charge is given
by  $n_{tot} = \sqrt{2/\pi} \; \partial_x \varphi_c$ plus finite
$q$ components. One can easily convince oneself that the
latter will give rise to subdominant contributions which
can be neglected in the scaling limit. Hence,
\begin{equation}
\begin{array}{rcl}
D (x) & = & \langle \cos \left[ \pi  \int_{0}^{x} dy
\left(  \sqrt{2/\pi} \; \partial_y \varphi (y) + \cdots \right) \right] \rangle
 \\
& \rightarrow & \langle \cos \left[ \sqrt{ 2 \pi} (\varphi ( x) - \varphi (0) )
\right] \rangle
 \\
& \sim & {1 \over { x^{K_c} } }
\end{array}
\label{Dbos}
\end{equation}

Since bosonization can only probe non-zero wave vector components
of the density the expressions are correct up to multiplicative
factors $\sim \cos( \pi \rho x)$ ($\rho$ is average density). Keeping 
this in mind, the outcome is fully consistent with the result
obtained for the large $U$ case (Eq. \ref{chstriSF}, $K_c = 1/2$ in this
limit) but now extended to arbitrary values of the charge stiffness!

The correspondence between bosonization
and the strong coupling analysis becomes very obvious 
in the derivations of the
two point spin correlator and the string correlator. Let us
recall the standard derivation in bosonization
of the spin correlator,
\begin{widetext}
\begin{equation}
\begin{array}{rcl}
 \langle S^z(x)S^z(0)\rangle 
&=& 
\displaystyle \frac{1}{2\pi}
\langle\frac{\partial \varphi_s(x)}{\partial x}
 \frac{\partial \varphi_s(0)}{\partial x} \rangle + 
\left[
\langle{\cal O}_{SDW,z}(x) {\cal O}_{SDW,z}^\dagger(0) \rangle+
{h.c.}\right] 
\vspace{1mm} \\
\langle{\cal O}_{SDW,z} (x){\cal O}_{SDW,z}^\dagger (0) 
\rangle& =& \displaystyle \frac{1}{8\pi^2} e^{-2ik_F x}
\langle e^{i \sqrt{2\pi} \left[ \varphi_c(x) - \varphi_c(0)\right]} \rangle
\langle e^{i \sqrt{2\pi} \left[ \varphi_s(x) - \varphi_s(0)\right]} \rangle
\vspace{1mm} \\
&=& \displaystyle \frac{1}{8\pi^2} e^{-2ik_F x} \frac{1}{x^{K_c + K_s}}
\label{bosp1}
\end{array}
\end{equation}
\end{widetext}
and  the spin-spin correlation function becomes
\begin{equation}
\langle S^z(x)S^z(0)\rangle =  -\frac{K_s}{4\pi^2} \frac{1}{x^2} +\frac{1}{4\pi^2}
\frac{\cos(2 k_F x)}{x^{K_c + K_s}}.
\label{bosp2}
\end{equation}

Comparing this with the large $U$ outcome, Eq.'s (\ref{SSlargeU}),
the correspondence is clear: $\langle e^{i \sqrt{2\pi} \left[ \varphi_s(x) 
- \varphi_s(0)\right]} \rangle$ is the staggered
magnetization of the spin chain in squeezed space, Eq. (\ref{Heiscor}). 
In strong coupling, the sublattice parity fluctuations enter via
the function $\langle n (-1)^{\sum n} n \rangle$ (Eq. \ref{chstrnSF})
which differs from
$D$ by just a factor $\cos( \pi \rho x)$. This is of course precisely
$e^{-2ik_F x}
\langle e^{i \sqrt{2\pi} \left[ \varphi_c(x) - \varphi_c(0)\right]} 
\rangle$ in the bosonization expression Eq. (\ref{bosp1}). Notice
that the subdominant uniform component $\sim 1 /x^2$ was just ignored
in the strong coupling analysis. 

The correspondence is further clarified by  considering the string correlator.
Straightforwardly,
\begin{widetext}
\begin{equation}
\begin{array}{rcl}
 \Ostr(x) &= & -
\langle S^z(x) (-1)^{\sum_{j=1}^{x-1} n_{tot}(j)} 
 S^z(0) \rangle \\
& = &
\left(
\frac{1}{4 \pi}
\langle
\partial_x  \varphi_s (x) 
 e^{  i  \sqrt{2 \pi}( \varphi_c(x) - \varphi_c(0) ) }
\partial_x  \varphi_s (0) \rangle + h.c. \right)
 \\
&&
\left(
\frac{e^{-2ik_F x}}{8 \pi^2}
\langle 
e^{i \sqrt{2 \pi} (\varphi_c(x) - \varphi_c(0) ) }
e^{-i \sqrt{2 \pi} (\varphi_c(x) - \varphi_c(0) ) }
\rangle
 \langle 
e^{-i \sqrt{2 \pi} (\varphi_s(x) - \varphi_s(0) ) }
\rangle + h.c. \right)
 \\
&&
\left(
\frac{e^{2ik_F x}}{8 \pi^2}
\langle 
e^{-i \sqrt{2 \pi} (\varphi_c(x) - \varphi_c(0) ) }
e^{-i \sqrt{2 \pi} (\varphi_c(x) - \varphi_c(0) ) }
\rangle
 \langle 
e^{-i \sqrt{2 \pi} (\varphi_s(x) - \varphi_s(0) ) }
\rangle + h.c. \right).  \\
\label{botop1}
\end{array}
\end{equation}
And these contributions add up to,
\begin{equation}
\begin{array}{rcl}
\Ostr(x) & = & -
\langle S^z(x) (-1)^{\sum_{j=1}^{x-1} n_{tot}(j)} 
 S^z(0) \rangle  \\
& = &
- \frac{1}{4 \pi^2} \frac{1}{x^{2+K_c}}
- \frac{1}{4 \pi^2} \frac{\cos(2k_F x)}{x^{K_s}}
- \frac{1}{4 \pi^2} \frac{\cos(2k_F x)}{x^{K_s+4K_c}}.
\label{botop2}
\end{array}
\end{equation}
\end{widetext}

The first term is obviously the (over corrected) uniform magnetization
and the leading singularity at finite wavevectors is,
\begin{equation}
\label{polkenm}
\langle S^z(x) (-1)^{\sum_{j=1}^{x-1} n_{tot}(j)} 
 S^z(0) \rangle = \frac{\cos(2k_F x)}{x^{K_s}}.
\label{botop3}
\end{equation}

Again the caveat applies that bosonization cannot keep track
of the average charge density  and the oscillatory factor in the
numerator should therefore be ignored -- this `flaw' is just inherited 
from $D(x)$, Eq. (\ref{Dbos}).
Where is this leading singularity coming from? It corresponds
with the third line in Eq. (\ref{botop1}). This algebra is expressing
that the charge vertex operator coming from the charge string exactly
compensates for the charge vertex operators attached to the spin operators.
We recognize that this is in precise correspondence with Eq.'s 
(\ref{pasotop}-\ref{toplUres}) of the strong coupling limit. 
The charge string is coding for the fluctuating kinks in the sublattice
parity and the string correlator is constructed to remove these from
the spin correlations.

What have we achieved? The above leaves no doubt that the algebraic
structure of bosonization
is exactly coding for the structure we discussed in
a geometrical language in section \ref{sec:luttinger}. However, in 
section \ref{sec:luttinger} we had
to rely on the simplifications arising in the strong coupling limit.
The algebraic structure of bosonization is however universal and 
independent of microscopic conditions like the strength of $U$.
For instance, in the non-interacting limit $K_c = K_s = 1$ and
one directly infers that the bosonization expressions Eq. (\ref{botop3},
\ref{bosp2}) are consistent with the exact results we derived 
for the string- and spin correlators for this limit in section \ref{sec:NonInteracting}.
Although there are some caveats regarding the use of bosonization
to calculate (charge) string correlators, these are entirely of a technical
nature and these affect only subdominant singularities: see appendix B.
We can therefore safely conclude that bosonization is just encoding the
squeezed space geometrical structure which is manifest in strong coupling.
The `hard-wired' structure of bosonization, in combination with the
string operators, leaves no room for any other conclusion that
squeezed spaces are ubiquitous in Luttinger liquids. It is indeed the
case that even  non-interacting one dimensional electron systems have deep
connections with hidden order in Heisenberg chains.

\section{Numerical Results}
\label{sec:numerical}

To verify that the correlator $\Ostr$ indeed demonstrates that 
squeezed space exists for finite values of the Hubbard coupling $U/t$
and arbitrary density, we performed numerical calculations using the
DMRG method\cite{white}. The DMRG is an ideal tool for these purposes,
because the algorithm construction implies that string correlators
are, in principle, no more difficult to construct than ordinary
two-point correlators. Indeed, the string operator $(-1)^{n_s}$ is
precisely that which is already used to ensure the correct commutation
relations for the creation and annihilation operators.
We utilized the non-abelian formulation\cite{nonabelian} of the DMRG, 
which makes use of the $SU(2) \otimes SU(2) \simeq SO(4)/Z_2$ spin and 
pseudo-spin symmetry of the Hubbard model\cite{so4},
thereby giving a substantial improvement in efficiency. The pseudo-spin 
symmetry is an expansion of $U(1)$ particle number symmetry $N$ to an 
$SU(2)$ symmetry which we denote here by $\vec{Q}$ (this is sometimes 
also denoted by $\vec{I}$). In the $SO(4)$ representation, the 
particle-number is given by the $z$-component of the pseudo-spin, 
$N = 2Q^z+1$. In our calculation, the basis states are $SO(4)$ multiplets, 
labeled by two half-integral quantum numbers $(s,q)$ denoting the total 
spin and total pseudo-spin respectively. 

Addressing the scaling limit with the DMRG method is subtle.
In the DMRG method, the ground-state wavefunction
is calculated in a Hilbert space which is truncated. The parameter 
controlling the truncation is the number of states kept in each `block', 
$m$. The actual dimension of the space in which the ground-state 
wavefunction is determined is of order $(4m)^2$. This truncation introduces
an error which, for a `well-behaved' system, is completely systematic and can be 
corrected for by calculating the appropriate scaling as 
$m \rightarrow \infty$. For the ground-state energy, this scaling is
understood and a routine calculation in DMRG. For correlation functions, 
the scaling is highly non-linear and difficult to perform, not least due to 
a result highlighted by \"Ostlund and Rommer\cite{rommer}:
the wavefunction obtained by DMRG is a (position-dependent) matrix-product 
wavefunction, which implies that the long-range asymptotic behavior of 
all two-point correlation functions is exponential,
with a correlation length that depends on the number of states kept $m$.
While in principle one can determine this correlation length and fit the 
remaining (algebraic) components of the correlation function, this is in 
fact not necessary due to a not so well understood property of 
(position-dependent) matrix product wavefunctions, namely in the 
short-distance correlations the exponential due to the finite truncation is 
not present at all. Thus, as long as a sufficiently large number of states are
kept to be close to the scaling limit at distances less than the
characteristic transition point where the correlator becomes exponential,
the exponents of algebraic terms can be determined with high accuracy
without any additional corrections due to the finite truncation.

Also of note is that matrix-product wavefunctions generically carry 
long-range string order, in the sense that it is likely that all string
correlation functions decay exponentially in the asymptotic limit, but it is 
permissible that
the decay is to a non-zero constant. The canonical example is the AKLT
wavefunction, which is obtained \textit{exactly} in (non-abelian) DMRG
with $m=1$ states kept. In principle, the variational nature of DMRG
implies that for a finite number of states kept one could inadvertently
and incorrectly obtain a state that has non-zero string order. 
This is not a serious issue and is 
entirely analogous to the case of ordinary two-point
correlators which, in the absence of a symmetry constraint, may have
a spurious (but usually negligible) non-decaying component.
For example, a not-quite-zero uniform
magnetization resulting in a non-zero constant
in the spin-spin correlator. The point is that
the construction of DMRG
treats hidden order of the den Nijs-Rommelse type on a very similar 
footing as more conventional order.

In the calculations presented here, we used $m=1000$ $SO(4)$ states kept, 
and a lattice size of $L=1000$. The lattice size was chosen to be rather 
large in an attempt to reduce the effect of the open boundary conditions.
However this is not strictly necessary and the usual averaging procedure
suffices to eliminate the Friedel oscillations and obtain the correct scaling 
form of the correlators even for much smaller lattices.

We calculated the string correlator $\Ostr$, Eq. (\ref{otop0}), 
the sublattice parity correlator $D$, Eq. (\ref{Ddef}) and
its second lattice-derivative $D_{nn}$, Eq. (\ref{Dnndef}), 
for a large variety of filling factors $\rho = 0.1 \ldots 0.9$ and 
$U/t = 0 \ldots 16$. Notice that the number operators appearing in
the `charge' strings $D$ and $D_{nn}$ correspond with $n_s$ measuring
the presence (1) or absence (0) of a singly occupied site. In the exponent
one might as well take the total charge density $n_{tot} = n_{\uparrow}
+ n_{\downarrow}$, i.e. $(-1)^{n_{tot}} = (-1)^{n_s}$. However,
$D_{nn} \sim \langle n_s \Pi (-1)^{n_s} n_s \rangle \neq       
\langle n_{tot} \Pi (-1)^{n_{tot}} n_{tot} \rangle$ because $n_s$ cannot
distinguish empty from doubly occupied sites whereas $n_{tot}$ does.
On the
bosonization level this subtlety does not matter, but it is consequential
for the numerically exact charge string correlators. As the strong coupling
analysis in section \ref{sec:NonInteracting} demonstrates, the charge string coding for the
squeezed space structure is actually $D_{nn}$ because empty and doubly
occupied sites are indistinguishable in the squeezing operation.

\begin{figure}[tbh]
\showgrace{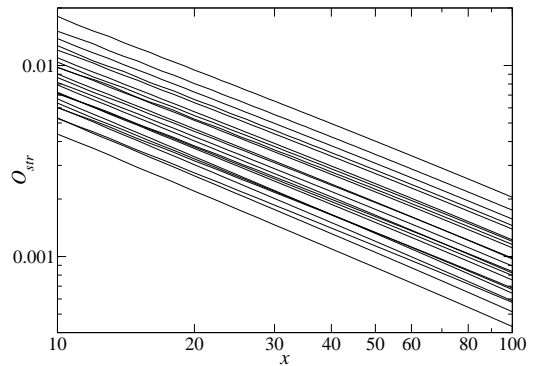}
\caption{The $\Ostr$ correlator on a log-log plot, for a wide variety of filling and couplings.
	The slope, which determines the exponent (up to log corrections) of the leading order term,
	is equal to 1 independent of the parameters.}
\label{fig:ses-loglog}
\end{figure}

The obtained correlation function $\Ostr$ appears in
Fig. (\ref{fig:ses-loglog}), plotted on a log-log scale. It is clear from the
figure that the leading order term in $\Ostr$ is algebraic, with an 
exponent that is independent of both the filling  factor and $U$. The fitted 
exponent is equal to 1, with a variation over all parameter ranges
of $\sim 5\%$. We percieve this as a striking result\cite{Letter}, taking away all
doubts regarding the `universality of squeezed space': regardless 
microscopic circumstances we have identified a correlation function
which always behaves as if the electron system is just the same spin-chain.

\begin{figure}[ht]
\showgrace{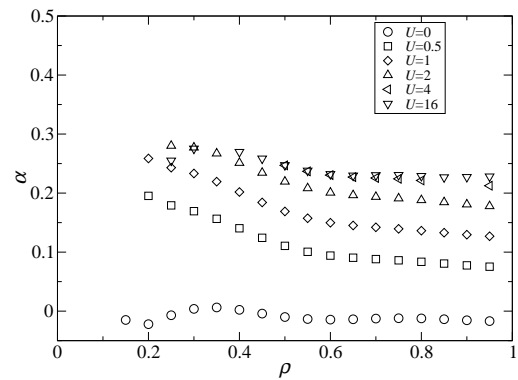}
\caption{The parameter $\alpha$ governing the logarithmic corrections to $\Ostr$. The logarithmic
	corrections are very sensitive to finite-size effects, which were not taken into account here.
	Thus the strong coupling limit deviates from the expected $\alpha = 1/2$.}
\label{fig:sps-logcorrection}
\end{figure}

Even the small variation of the exponent  is explainable, employing 
logarithmic  corrections. 
At the $U/t \rightarrow \infty$ Woynarovich-Ogata-Shiba point, the 
wavefunction factorizes exactly and the $\Ostr$ correlator measures 
exactly the logarithmic corrections of the isotropic $S=1/2$ 
antiferromagnetic Heisenberg chain\cite{affleckheis,Singh}.
This coincides with the well-known form at half-filling\cite{finkelstein}, 
where the presence of the charge gap implies Heisenberg-like behavior of 
the logarithmic corrections for any $U > 0$. However, as far as we know, for 
finite $U/t$ away from half-filling, the exact form of the logarithmic 
corrections is not known. It is well understood that these corrections
are not governed by the conformal field theory, implying that these are
non-universal quantities which are sensitive to the short distance dynamics.
Hence, they are not necessarily independent of doping and interaction
strength. This is confirmed by the non-interacting limit $U =0$ where the 
system has no log corrections, and there is no reason to expect a 
discontinuity between $U = 0$ and $U \rightarrow \infty$. Therefore we take 
the scaling  form of the string correlator to be
\begin{equation}
\Ostr (x)  = A (\rho, U) \frac{\ln^\alpha (x)}{x}
\label{eq:logcorrection}
\end{equation}
and fit for the running exponent $\alpha$. The obtained exponent appears in 
Fig. (\ref{fig:sps-logcorrection}). We emphasize that this is a very rough 
calculation, obtained by a direct fit of the finite-size data to the 
asymptotic form, ignoring finite-size corrections.
For the Heisenberg model these corrections are important\cite{hallberg} therefore 
our $U \rightarrow \infty$ estimate of $\alpha \sim 0.25 - 0.3$, 
deviates from the exact value of $\alpha = 1/2$. Indeed, our result is
rather reminiscent of earlier Heisenberg model calculations\cite{scalapino}
which suffer from similar issues.
A careful scaling analysis, done by Hallberg, Horsch and Martinez 
for the Heisenberg chain\cite{hallberg}, 
should  present no difficulty and will be reported in a subsequent 
paper. However, from the present results we can already safely
conclude that $\alpha$ is a function of $U/t$ and perhaps also density.

\begin{figure}[ht]
\showgrace{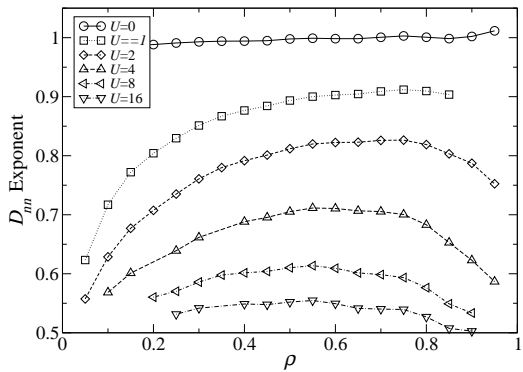}
\caption{The exponent of $D_{nn}(x)$. This function isolates the charge contribution to the
	correlation functions, hence gives a direct determination of $\Kc$. The solid lines are guides to the eye.}
\label{fig:dnn-exponent}
\end{figure}

We have argued in previous sections that the charge fluctuations present 
in the ordinary two-point correlators are due to sublattice parity 
fluctuations. We found in section \ref{sec:luttinger} that in the strong coupling limit
the following rigorous result holds for the staggered component of
the spin-spin correlator,
\begin{equation}
\langle \vec{S}(0) \vec{S}(x)\rangle \sim \Ostr(x) D_{nn}(x) \; .
\label{SSrep}
\end{equation}
Our argument is that bosonization reflects this
structure and we are now in the position to test this relation
numerically for arbitrary values of $U$ and density. As we already
emphasized, to isolate the squeezed space the number operators in
$D_{nn}$ should measure the density of singly occupied sites, $n_s$. In 
addition, away from the Woynarovich-Ogata-Shiba point Eq. (\ref{SSrep})
is not longer exact but it should become exact in the scaling limit.
Eq. (\ref{SSrep}) should hold up to a $U, \rho$ dependent prefactor
factor which is set by short distance physics. This is exactly what we
find. This is demonstrated by Fig. (\ref{fig:dnn-exponent}) which shows the 
exponent of the $D_{nn}$ correlator, which turns out to be given by
  \begin{equation}
D_{nn}(x) = B (\rho, U)  \frac{cos(2k_F x)}{x^\Kc} + O(x^{-1-\Kc}) \; .
\label{Dnnnum}
\end{equation}
where $\Kc$ is the usual density and $U$-dependent charge stiffness
of the Hubbard model. It follows that 
\begin{equation}
\langle \vec{S}(0) \vec{S}(x)\rangle = F (\rho, U) 
\frac{cos(2k_F x)}{x^{\Kc + 1}}
\ln^\alpha(x)
\label{SSnum}
\end{equation}
coincident with the well known asymptotic behavior of the two point
spin correlator in the Luttinger liquid. This completes our case.
The fact that we not only isolate the spin-only dynamics in
the Luttinger liquid using $\Ostr$ but that we can reconstruct
the two point spin correlator by dressing it with an entity
which is exclusively counting the sublattice parity mismatches
($D_{nn}$) leaves no doubt that squeezed space is universal.

Let us end this section with giving some numerical results 
regarding the non universal prefactors $A(\rho, U)$, $B(\rho, U)$
and $F (\rho, U)$. These are clearly sensitive to the details 
of the short wavelength dynamics and have therefore a similar status as
non-universal amplitudes in any critical theory. Hence, these
have to be calculated numerically.   

\begin{figure}[ht]
\showgrace{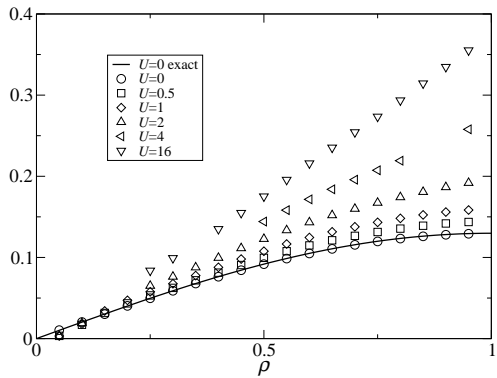}
\caption{The prefactor of $\Ostr$. The numerical data at $U=0$ matches the exact form determined in section 
\protect{\ref{sec:NonInteracting}}. The $U \rightarrow \infty$ prefactor is proportional to the density,
exactly as required for a diluted Heisenberg chain. }
\label{fig:ses-prefactor-with-log-correction}
\end{figure}

The prefactor of the $\Ostr$ string correlator, $A(\rho,U)$ is given in figure \ref{fig:ses-prefactor-with-log-correction}.
The numerical prefactor coincides with the expected exact expression at $U=0$ and follows the expected
form $\propto \rho$ for $U \rightarrow \infty$, for a Heisenberg chain diluted by a 
hole density of $(1-\rho)$. The exact slope of the $U \rightarrow \infty$ prefactor depends sensitively
on the exponent of the log corrections. The underestimation of $\alpha$ in equation (\ref{eq:logcorrection})
results in the prefactor of figure \ref{fig:ses-prefactor-with-log-correction} being somewhat large;
the $U \rightarrow \infty$ form should be\cite{affleck}
\begin{equation}
\Ostr(x) = \frac{3}{(2\pi)^{3/2}} \frac{\rho \ln^{1/2}(\rho x)}{x} \; .
\end{equation}

This differs from the correlator of a stretched Heisenberg chain by a prefactor $\rho^2$, 
which is due to the dilution of the spins; for the Heisenberg chain $\langle s \rangle = 1/2$,
but for the Hubbard model $\langle s \rangle = n_s / 2$. Thus, with all prefactors accounted for,
the factorization of the $U \rightarrow \infty$ spin correlator is\cite{Parola,affleck}
\begin{widetext}
\begin{equation}
\begin{array}{rcl}
\langle \vec{S}(0) \vec{S}(x) \rangle & = & \displaystyle -\frac{3}{4(\pi x)^2} + \frac{1}{\rho^2} \Ostr(x) D_{nn}(x) \\
	                                          & = & \displaystyle -\frac{3}{4(\pi x)^2} + 
	\frac{3 A^2}{(2 \pi)^{3/2}} \frac{\cos(2k_F)}{\rho \sqrt{\sin(2k_F)}}
	\frac{\cos(2k_F x) \ln^{1/2}(x)}{x^{3/2}} \; ,
\end{array}
\label{eq:ss-decomposition}
\end{equation}
\end{widetext}
with $2k_F = \pi \rho$.

\begin{figure}[ht]
\showgrace{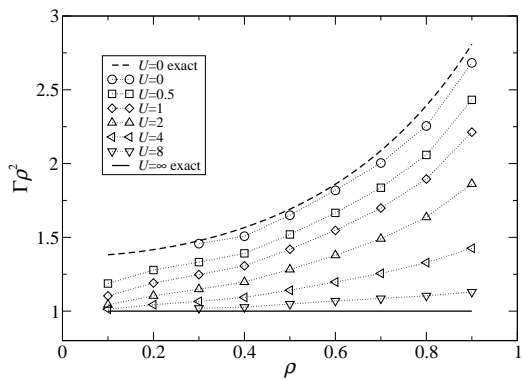}
\caption{The aplitude $\Gamma$ obtained from
$\langle \vec{S}(0) \vec{S}(x) \rangle \sim \Gamma \Ostr(x) D_{nn}(x)$. 
For clarity plotted as $\Gamma \rho^2$, which
is equal to unity in the strong coupling limit. The heavy solid and dashed lines are the exact expressions
at $U = \infty$ and $U = 0$ respectively. The light dashed lines linking the numerical data are guides for the eye. }
\label{fig:gamma}
\end{figure}

For finite coupling the exact factorization of the wavefunction is destroyed by local fluctations, so
Eq. (\ref{eq:ss-decomposition}) only applies rigorously in the strong coupling limit. 
As shown in section \ref{sec:NonInteracting} however, the
scaling form applies even to $U=0$, with the introduction of a non-universal amplitude $\Gamma(U,\rho)$,
\begin{equation}
\langle \vec{S}(0) \vec{S}(x) \rangle = -\frac{3}{4 (\pi x)^2} + \Gamma(U,\rho) \, \Ostr(x) \, D_{nn}(x) \; .
\end{equation}
Figure \ref{fig:gamma} shows this amplitude as a function of density and $U$, which is always finite implying 
that squeezed space is ubiquitous.

\section{Conclusions: the fermion minus signs.}
\label{sec:conclusion}

In first instance the pursuit presented above can be seen
as an exploration of the usefulness of string correlators of the
den Nijs and Rommelse type in the context of one dimensional
physics. To our perception these correlation functions  are worthy 
additions to the standard repertoire of one dimensional physics. This
will be further amplified in a next paper where we will further 
explore the information one can obtain from string correlators
like $D$ and $D_{nn}$.

In this paper we used string correlators to clarify some conceptual
issues in one dimensional physics. String correlators go hand in hand
with the simple geometrical ideas which emerged in the study of Haldane
spin chains and the strong coupling Bethe Ansatz solution of the Hubbard
model. These correlators make it possible to address to what extent
these notions are of relevance to generic Luttinger liquids and we
made the case that squeezed spaces are hard-wired into Luttinger liquid
theory. It is merely a matter of recognition. 

Although complementary to the standard descriptions, we find that 
the squeezed space notion does exert unifying influences.   
It is not an accident that we started out discussing the Haldane spin chains.
We hope that we convinced the reader that there is a unity underneath
which becomes obvious in this language, while it is far from obvious
in the standard formulation of bosonization.

Is it more than just clarification? If so, it should be that these 
insights can be used to deduce states of one dimensional quantum
matter which have been overlooked before. In the Luttinger liquid
context we have deduced one such novel state: the `charge only'
superconductor we introduced at the end of section \ref{sec:luttinger}. This
entity can also be discussed in the bosonization language. It is
a prerequisite to drive the system away from critically such that
the charge sector is genuinely disordered. This requires an external 
Josephson field stabilizing superfluid phase order. A conventional
Josephson field acting on electrons pairs in the singlet channel
is expressed as (recall section \ref{subsec:dictionary}),
\begin{equation}
\begin{array}{rcl}
H_J & = & B_J \int dx \left( \psi_{\uparrow} (x) \bar{\psi}_{\downarrow} (x)
- \psi_{\downarrow} (x) \bar{\psi}_{\uparrow} (x) \right)
 \\
      & \sim & B_J \int dx \cos \left[ \sqrt{ 2 \pi} \theta_c \right]
\; \sin \left[ \sqrt{ 2 \pi} \varphi_s  \right]
\end{array}
\label{josbos}
\end{equation}
involving the dual charge field  $\partial_x \theta_c (x) = -\Pi (x)$.
This imposes phase order (pinning of $\theta_c$) but it has also the
immediate effect of opening a spin gap ($\sim \sin \left[ \sqrt{ 2 \pi} 
\varphi_s  \right]$). This spin gap means that the spins are paired in
pairwise singlets and a squeezed space cannot be defined for these
singlets. Instead, what is required is a Josephson field acting 
exclusively on the charge fields,
\begin{equation}
H'_J =  B_J \int dx \cos \left[ \sqrt{ 2 \pi} \theta_c \right]
\label{josboc}
\end{equation}
This will enforce disorder on the charge sector, leaving the
spin sector unaffected. Recalling the discussion of the spin chain,
this charge disorder turns into a $Z_2$ gauge invariance in the
spin sector. The spin system in squeezed space resides at the
($SU(2)$) critical point separating the XY and Ising fixed 
points and together with the minimal coupling to the deconfining
$Z_2$ gauge fields a state of matter is realized which is 
symmetry-wise indistinguishable from the critical state of the
Haldane spin chain found at the transition from the hidden-order phase
to the $S=1$ XY phase. 

Although such a state is a theoretical possibility, it is less
clear whether it can be realized in nature. Bosonization is helpful
in clarifying this issue. Starting out with electron operators,
it appears to be impossible to construct a Josephson field of the
form Eq. (\ref{josboc}). One will always find that the charge
Josephson field is accompanied by a (relevant) operator in
the spin sector. This might well turn out to be a fundamental
obstruction. In the one dimensional universe the charge and
spin fields are more fundamental than electrons, and a-priori
Eq. (\ref{josboc}) is physical. However, a Josephson field will
in practice correspond with a mean-field coming from three
dimensional interactions and this implies that this mean-field
has to be a composite of electron degrees of freedom.    

As we argued, squeezed space is hard-wired into the bosonization
formalism and even exotic states like those discussed in the
previous paragraphs are in principle within the reach of the formalism.
By implication, if a state of electron matter would exist where
squeezed space is destroyed, it would be beyond bosonization. In
the context of the (bosonic) spin matter of the Haldane chain we
encountered this possibility. Helped by the identification of
the $Z_2$ gauge symmetry, we presented a recipe (the transversal
field) to stabilize a non-squeezed space (`confining') phase of
the spin chain. Is this also possible in the electron Luttinger 
liquids?

In this regard it is helpful to view these matters from a yet 
another angle: the Marshall signs introduced by Weng in the
one dimensional context\cite{Weng} as an addition to the squeezed
space construction needed to describe fermion propagators; see
also reference \onlinecite{Weng1} for the extension to 2D and for some interesting
observations regarding Marshall signs and spin-charge separation
in 1D. Marshall signs refer to the
theorem that the ground state wave function of a $S=1/2$ spin system
defined on a bipartite lattice with nearest neighbor exchange 
interactions is nodeless: it is a bosonic state. In the strong coupling
limit the spin system in squeezed space is of this kind, and this
explains in turn why the Bethe-Ansatz solution reveals that the
charges are governed by spinless fermions. The total wavefunction has
to be anti-symmetric and because squeezed space exists the spin sector
is symmetric, so that the fermionic grading resides in the charge sector.

Although we are not aware of an explicit proof, it has to be that this
`division of statistics' is universal in the scaling limit. Our
string correlator demonstrates that at  long distances the squeezed
space spin system does behave exactly like the (unfrustrated) Heisenberg
chain and it is hard to imagine that this would survive a drastic change
involving the nodal structure of the spin wavefunction. Let us assume
that the strong coupling limit is in this regard a prototype of any
Luttinger liquid, to recollect the lessons learned from the bosonic
spin chain. There we learned that to break up squeezed space `charge'
fluctuations are needed  changing its length from odd to even and 
vice versa. This implies that single charges can be created or annihilated 
and this is of course not a problem in a bosonic system because a single
boson can condense. However, single fermions cannot condense and since
in the Luttinger liquid for reasons just discussed the charge sector
is fermionic, confinement is impossible. Admittedly, the argument is
circular. It starts out postulating the existence of squeezed
space as an entity unfrustrating the spin system in the Marshall sign
sense, to find out that the minus signs in turn offer a complete
protection of the squeezed space. This viewpoint suggests
that there might be ways around the squeezed space and that states
can be constructed which are beyond bosonization. Starting from
strongly coupled microscopic dynamics, one can image interactions 
which are strongly frustrating the spin system in the Marshall sign
sense (i.e. longer range spin-spin interactions). Such interactions
could lead to a `signful' spin physics in squeezed space, which
in turn could diminish the `statistical protection', possibly leading
to metallic states which are not Luttinger liquids. 

A final issue is, is there anything to be learned regarding the
relevance of Luttinger liquid physics in higher dimensions?
In this paper we have worked hard to persuade the reader that
squeezed space is a {\em defining} property of the Luttinger
liquid. As such, it is a-priori not special to one dimension,
in contrast to e.g. the lines of critical points and the Mandelstam
construction. Given a complete freedom to choose the
microscopic conditions, which fundamental requirements should be
fulfilled to form squeezed spaces in higher dimensions? First,
bipartiteness is required and this is no longer automatic
in higher dimensions. As a starting point one needs a Mott-insulator
living on a bipartite lattice characterized by an unfrustrated,
colinear antiferromagnet. Upon doping such a Mott-insulator the
charges (holes) will frustrate this spin system unless special 
conditions are fulfilled: these holes have to form $\rm D-1$ dimensional
connected manifolds as a fundamental requirement to end up in
a bipartite space after the squeezing operation. Different from
the one dimensional situation, true long range order will take over
when it gets a chance. A first possibility is that these
$\rm D-1$ dimensional hole manifolds simply crystallize, forming 
charge ordered state accompanied by a spin system showing a
strong ordering tendency  as well, with the characteristic that the
staggered order parameter flips every time a charge-manifold is
crossed. One immediately recognizes the stripe phases which are 
experimentally observed in a variety of quasi-2D Mott-insulators, including
the cuprates\cite{Zaanenphilmag}. 
Alternatively, assuming that the holes move in pairs,
general reasons are available demonstrating that the charge sector
can turn into a superconductor (via a dual dislocation 
condensation\cite{qunematic}) 
such that the manifolds continue to form domain walls in the sublattice
parity although their locus in space is indeterminate. In direct analogy 
with the Haldane spin chain, such a state is characterized by an
emergent `sublattice parity' $Z_2$ gauge invariance. 

The above is just a short summary of some aspects of the `stripe 
fractionalization' ideas and for a further discussion we 
refer to the literature\cite{Zaanenphilmag,Nussinov,Sachdev0,Sachdev1}.
Most importantly, the notion of squeezed 
space make it clear why `Luttinger liquid-like' physics is not at all 
generic in higher dimensions but instead rather fragile, if it exists 
at all. The bipartiteness of squeezed space-time in the space directions 
has to be protected and this requires microscopic fine-tuning.   

The punchline is that if one wants to contemplate  manifestations
of Luttinger liquid physics in higher dimensions it \textit{must} be
striped in one way or the other, since squeezed spaces are 
the most precise way to characterize the 
phenomenon of spin-charge separation as it arises in the specific
one dimensional context. This insight also makes is clear why attempts
to invoke the {\em equations} governing the Luttinger liquids in
whatever phenomenological spirit to explain physics in higher 
dimensions are bound to fail: these represent a dynamics which
is slaved to  an underlying geometrical principle which is only 
of the right kind in one space dimension. To bosonize the electron
itself in two space dimensions one has to invoke geometrical/gauge
principles of a fundamentally different kind\cite{Fradkin,Weng1}.  
  
\begin{acknowledgments}
This work profited much from discussions
with S. Sachdev, S.A. Kivelson, Z.Y. Weng, G. Ortiz, C.D. Batista,
B. Leurs, F. Wilczek and M. Gul\'acsi. The work was supported by 
the Netherlands Foundation for Fundamental research of Matter 
(FOM). Numerical calculations were performed at the 
National Facility of the 
Australian Partnership for Advanced Computing, via a grant from
the Australian National University Supercomputer Time Allocation Committee.
\end{acknowledgments}

\appendix

\section{Computation of the charge string correlator of free
spinless fermions.}
\label{app:string}
    
In this appendix we discuss the numerical computation of
the free spinless fermion charge string operator, Eq.(\ref{Ddef}). 
We find that it can be fitted very accurately with the simple 
expression Eq. (\ref{chstriSF}). This may well be an exact result
but we did not manage to find the solution with analytical means.

Using periodic boundary conditions,
the charge string correlator can be written as
\begin{widetext}
\begin{equation}
\begin{array}{rcl}
\langle \,  (-1)^{ \sum_{j=1}^{x-1}
   n (j)}  \rangle &=&
 \langle k_N \ldots k_1 | (-1)^{ \sum_{j=1}^{x-1}
   n_{SF}(j)   }\, | k_1 \ldots k_N \rangle_{SF}   \\ 
&=&
 \sum_{x_1 \ldots x_N}
\sum_{y_1 \ldots y_N} 
\langle 
0|
a_{x_N} \ldots a_{x_1}  (-1)^{\sum_{j=1}^{x-1} n (j) }
  a_{y_1}^\dagger \ldots a_{y_N}^\dagger  
|0
\rangle
 \\
&& \times
\left(\frac{1}{V}\right)^N
e^{- ik_1 x_1 - \ldots - ik_N x_N}
~
e^{ ik_1 y_1 + \ldots + ik_N y_N} 
  \\
&=&
 \sum_{x_1 \ldots x_N}
\sum_{y_1 \ldots y_N} 
\langle 
0|
a_{x_N} \ldots a_{x_1} 
  a_{y_1}^\dagger \ldots a_{y_N}^\dagger  
|0
\rangle
 \\
&&
 \times
\left(\frac{1}{V}\right)^N
e^{- ik_1 x_1 - \ldots - ik_N x_N}
~ e^{ ik_1 y_1 + \ldots + ik_N y_N}  \\
&& \times
\prod_{j=1}^N \left[ 1- 2 \theta(y_j -1) \theta (r - 1 - y_j)  \right] \; .
\end{array}
\label{head12}
\end{equation}
occupying the lowest N $|k_1 \ldots k_N \rangle$ single fermion states.
The product term on the last line equals $-1$ when $y_j \in
 \left[ 1, x-1 \right]$ and 1 otherwise, taking into account the result of the
factor $(-1)^{\sum_{j=1}^{x-1} n_{SF}(j)} $.
Part of this sum can be written as
\begin{eqnarray}
\frac{1}{V} \sum_{y} e^{iy(p-k)} 
\left[ 1- 2 \theta(y -1) \theta (x - 1 - y)  \right]  &=& 
 \frac{1}{V} \sum_{y} e^{iy(p-k)}  - \frac{2}{V} 
\sum_{y =1}^{x-1} e^{iy (p-k)} \nonumber \\
&=& \delta(p,k) - \frac{2}{V} 
\frac{ e^{i(p-k) x }  - e^{i(p-k)}  }{ e^{i(p-k) }-1}
\nonumber \\
&\equiv& \delta^*(p,k),
\end{eqnarray}
abbreviating the second line with the  `star-delta function' $\delta^*(p,k) $.
Using this function, 
the expression (\ref{head12})
can be expressed as the determinant of a $N \times N$
 matrix  containing $\delta^*(k_i, k_j) $ functions
\begin{equation}
\label{term1}
\langle (-1)^{\sum_{j=1}^{x-1} n_{SF}(j)} \rangle= 
\mbox{det}
\left(
\begin{array}{cccc}
\delta^*(k_1 , k_1) & \delta^*(k_2 , k_1) & \cdots& \delta^*(k_N , k_1)  \\
\delta^*(k_1 , k_2) & \delta^*(k_2 , k_2) & \cdots& \delta^*(k_N , k_2)  \\
\delta^*(k_1 , k_3) & \delta^*(k_2 , k_3) & \cdots& \delta^*(k_N , k_3)  \\
 \vdots & \vdots & \vdots & \vdots\\
\delta^*(k_1 , k_N) & \delta^*(k_2 , k_N) & \cdots & \delta^*(k_N , k_N)  \\
\end{array}
\right).
\end{equation}
and this determinant can be straightforwardly computed numerically
for a finite system.
\end{widetext}

Careful analysis of the numerical data for a complete range of densities, 
 demonstrates that,
\begin{equation}
\langle \, (-1)^{\sum_{j=1}^{x-1} n (j) } \,
\rangle_{SF} 
= 
 \frac{A^2 \sqrt{2}}{\sqrt{\sin(\frac{\pi N}{V})}}
 \frac{ \cos(\frac{\pi (x-1) N}{V}) }{\sqrt{ 
 \frac{V}{\pi} \sin(\frac{\pi( x-1)}{V})
}
} \; .
\label{prompt}
\end{equation}

\begin{figure}[tbh]
\includegraphics[angle=270,width=\figurewidth]{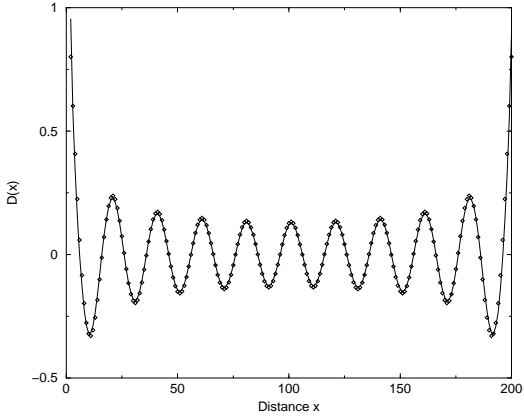}
\caption{Numerical results (circles) for the function
$D(x) = \langle (-1)^{\sum_{j=0}^x n_{SF}(j)}\rangle$
calculated from (\ref{term1}), as compared 
to the analytical form  Eq. (\ref{prompt}) (full line)
This is a representative example: we use $N = 20$ particles
on  on a chain of length $V=200$ (density  $\rho_{SF}= \frac{N}{V} =  0.1$),
using periodic boundary conditions.}
 density $\rho_{SF}= \frac{N}{V} =  0.1$.
\label{Dxvoor20}
\end{figure}

\begin{figure}[tbh]
\includegraphics[angle=270,width=\figurewidth]{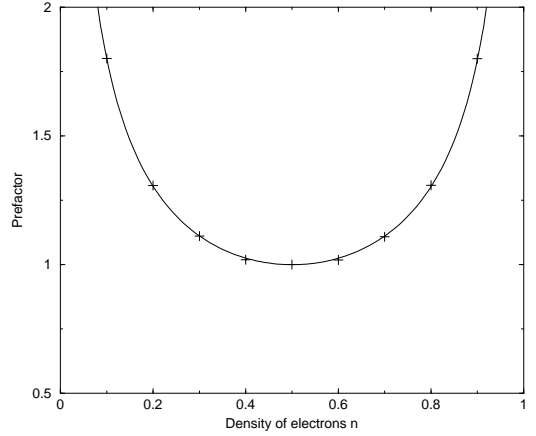}
\caption{The crosses indicate the numerical results for the 
prefactor of the function 
$D(x) = \langle (-1)^{\sum_{j=0}^x n_{SF}(j) } \rangle$
normalized to 1 for $n = \rho_{SF}= \frac{N}{V} = 0.5$ ($V=200$). 
The full line corresponds with the
function $\frac{1}{\sqrt{ \sin(\pi \rho_{SF})}}$.}
\label{Dxopeen}
\end{figure}

As an example, in  Fig. (\ref{Dxvoor20}) we show results for
$\rho_{SF} = \frac{N}{V} = 0.1$ for
$N=20$ particles on a chain of length  $V=200$ and this compared with
the  analytic expression (\ref{prompt}).
In Fig. (\ref{Dxopeen}) the numerical outcomes for the
prefactor of $\langle (-1)^{\sum_{j=1}^{x-1}n_{SF}(j)} \rangle$ taking
a normalization such that this prefactor is
 1 for $\rho_{SF}= \frac{N}{V}=0.5$. According to the exact result by 
Parola and Sorella\cite{Parola} this prefactor is equal to
$\langle (-1)^{\sum_{j=1}^{x-1} n_{SF}(j)}\rangle
 \sqrt{\frac{V}{\pi} \sin (\frac{\pi (x-1)}{V}) } / (A^2 \sqrt{2} 
\cos(\frac{\pi (x-1) N}{V}))$. The perfect match between this
normalized  numerical outcome and the function 
$\frac{1}{\sqrt{\sin (\pi \rho_{SF})}}$ establishes the density 
dependence of the amplitude in Eq. (\ref{prompt}).

In the thermodynamic limit $V\rightarrow \infty, 
\frac{N}{V} \rightarrow \rho_{SF}$
Eq. (\ref{prompt}) becomes,
 \begin{equation}
\langle \, (-1)^{\sum_{j=1}^{x-1} n (j) } \,
\rangle_{SF} =
 \frac{A^2 \sqrt{2}}{\sqrt{\sin( \pi \rho_{SF} )}}
 \frac{ \cos( \pi \rho_{SF} (x-1) ) }{\sqrt{x-1}
}.
\label{perlier}
\end{equation}
reproducing the exact result by Parola and Sorella\cite{Parola} at the density
$\rho_{SF} = \frac{1}{2}$. These authors  showed that at this specific density
the asymptotic form of $D(x)$ is
\begin{equation}
D(x)= \langle \, (-1)^{\sum_{j=0}^x n_{SF}(j) } \,
\rangle
= A^2\sqrt{2}
 \frac{\cos(\frac{\pi (x+1)}{2})}{ \sqrt{x+1} },
\end{equation}

For completeness, let us list the outcomes for the expectation
values  
$\langle n (x) (-1)^{ \sum_{j=1}^{x-1}
   n (j)}  \rangle_SF $  
and $\langle \,  n (x)   (-1)^{ \sum_{j=1}^{x-1}
   n (j)} n (0)   \rangle$ which can be regarded as lattice
derivatives of the charge string correlator Eq. (\ref{perlier}).
Eg find
\begin{widetext}
\begin{equation}
\begin{array}{rcl}
\lefteqn{\langle \, n (x)\, (-1)^{   \sum_{j=1}^{x-1} n (j)   }
\,\rangle_{SF} \, }~~~ \nonumber  \\
&=&
\langle \,(-1)^{  \sum_{j=1}^{x-1} n (j)  }
\,n_{SF}(0) \,\rangle_{SF} \,
=
\frac{D(x-2) - D(x-1)}{2} \nonumber \\
&= &
\frac{A^2}{ \sqrt{2 \sin( \pi \rho_{SF} )}}
\left(
\frac{ \cos( \pi \rho_{SF} x) \left[ \cos (\pi \rho_{SF}) -1 \right]
+ \sin(\pi \rho_{SF} x ) \sin (\pi \rho_{SF})
 }{    \sqrt{x } } \right) \nonumber \\
&=&\mbox{sign} \left[\cos (\pi \rho_{SF}) -1 \right]
 \frac{A^2 \sqrt{1 - \cos(\pi \rho_{SF})}}{ \sqrt{\sin(\pi \rho_{SF})}} 
\frac{\cos( \pi \rho_{SF} x  -K)}{\sqrt{
x } } \; , 
\end{array}
\label{option}
\end{equation}
where the constant $K$ is given by
\begin{equation}
K = \frac{\pi(\rho-1)}{2} \; .
\end{equation}
In addition,
\begin{equation}
\begin{array}{rcl}
\langle \,n (x) \, (-1)^{ \sum_{j=1}^{x-1}  n (j) }\, n (0)
\, \rangle_{SF}
& = & \displaystyle \frac{1}{4}
\left[ D(x-2) - 2D(x-1) + D(x) \right] \vspace{1mm} \\
&=& \displaystyle
\frac{A^2 (\cos(\pi \rho_{SF})-1 )}{ \sqrt{2 \sin(\pi \rho_{SF})}}
\frac{\cos(\pi \rho_{SF} x)}{ \sqrt{
x
 }} \; .
\end{array}
\label{moin}
\end{equation}
\end{widetext}

\section{A caveat: Schulz' flawed logic.}
\label{app:schulz}

The idea that the charge fluctuates the space in which the spin
system resides has a  long history. In this
appendix we would like to comment on an  argument due to the late 
Heinz Schulz\cite{Schulz1}. His argument is not correct, but the flaw is 
subtle and
informative regarding the workings of sublattice parity fluctuations.

In the above, we re-derived the `classic'  result that the two point
spin correlator $\langle S S \rangle \sim 1 / x^{K_c + K_s}$. 
Schulz\cite{Schulz1} asserted that this behavior can be explained
by assuming that the system can be seen as a 1+1D harmonic crystal
of charges in the continuum. The spins at the sites of this crystal 
would just  form a Heisenberg antiferromagnet. True
long range crystal order is impossible in 1+1D because the admixture
of the Goldstone bosons (phonons) renders the correlations to be
algebraic (algebraic long range order, ALRO). Schulz' idea was 
simple: the spin systems does not live on fixed positions in 
space but instead on a medium undergoing gaussian fluctuations,
as if the spin system `surfs' on the gaussian charge waves. 

The effects on the spin correlator can be easily calculated.
In the continuum  the spin density equals
\begin{equation}
\vec{S}(x) = \sum_m \vec{S}_{Heis.}(m) \delta(x-x_m),
\end{equation}
summing over all the electrons. Starting from the ALRO crystal, $x_m$
can be written as $x_m = R_m + u_m$,  where  $R_m = \frac{m}{\rho}$
 is the position  in the
$m$-th electron and $u_m$ its displacement. One finds for the
correlation function,
\begin{widetext}
\begin{equation}
\langle \vec{S}(x) \vec{S} (0)\rangle  \sim  \int d q  
 \sum_{m,m'} 
e^{-iqx} \langle \vec{S}_{Heis.}(m)
\vec{S}_{Heis.}(m')\rangle e^{iq(R_m -R_{m'})} 
\langle e^{iq( u_m - u_{m'})}\rangle \; .
\label{Schulz1}
\end{equation}
Due to the gaussian fluctuations, 
\begin{equation}
\langle e^{iq( u_m - u_{m'})}\rangle  \approx |m-m'|^{- \alpha(q)},
\end{equation}
with $\alpha(q) \sim q^2$. 
The $q$ integration in Eq. (\ref{Schulz1}) is dominated by the term
 $q \approx \pi \rho = 2k_F$ and  
using the Heisenberg correlation function, Eq.  (\ref{Heiscor}),
\begin{eqnarray}
\label{label6}
\langle \vec{S}(x) \vec{S}(0)\rangle  & \approx&  \int d q 
 \sum_{m,m'} e^{iq(R_m -R_{m'} -x)}   \langle \vec{S}_{Heis.}(m)
 \vec{S}_{Heis.}(m')\rangle 
|m-m'|^{-\alpha(2k_F)} \nonumber \\
& \approx&  \int d q 
 \sum_{m,m'} e^{iq(\frac{m}{\rho} - \frac{m'}{\rho} -x)}
  \frac{(-1)^{m-m'}}{
|m-m'|^{1+\alpha(2k_F)}} 
\nonumber \\
&=&
\frac{\cos(\pi \rho x ) \ln^{1/2} (\rho x)}{(\rho x)^{1+ 
\alpha(2 k_F)}} =
 \frac{\cos(2k_F x ) }{(\rho x)^{1+ \alpha(2k_F)}} \ln^{1/2} (\rho x).
\label{schulz2}
\end{eqnarray}
\end{widetext}

This outcome looks indeed quite like the desired result, identifying
$\alpha (2k_F)$ with $K_c$ . However, this similarity is actually 
misleading! Schulz' crystal refers to the breaking of translation
symmetry by single electron charges. Implicitly, this refers to the
strongly coupled regime considered in the above and this crystal 
corresponds with  the spinless-fermion ALRO crystal (e.g. reference \onlinecite{Zaanenstr}).
The spinless fermion $2k_F$ turns into a spinful electron $4k_F$ wave
vector. Accordingly, the exponent  $\alpha (2k_F)$ should be associated
with the charge stiffness appearing in the $4k_F$ charge correlations,
and this stiffness is not $K_c$ but instead $4K_c$! For instance, in the
large $U$ case $K_c = 1/2$ and the Schulz argument would predict
that the spin correlations would decay like $1/x^3$ instead of $1/x^{3/2}$.   

Where is the flaw? In fact, the implicit assertion in the above is that 
$\langle S S \rangle \sim \langle \; n \;  n \; \rangle_{4k_F} 
\times \langle
S S \rangle_{Heis.}$. We learned, however, that the geometry of
the spin system is fluctuated by {\em kinks} in their translational
sector (the sublattice parity flips). These are dual to the charge-order
and one has to use instead the exponentiated charge strings
$\langle S S \rangle \sim \langle n (-1)^{\sum n} n \rangle    
\times \langle S S \rangle_{Heis.}$.  As we showed, $\langle n (-1)^{\sum n} n 
\rangle$ decays with an exponent which is $K_c$ itself. 
From the discussion in section \ref{sec:NonInteracting} it is clear
that this dual structure is in fact respected by 
bosonization. In this sense, bosonization `knows' about squeezed space.

\end{document}